\documentstyle[emulateapj,psfig,url]{article}

\newcommand{\lya}{Ly$\alpha$}

\makeatletter

\newenvironment{inlinefigure}{%
\def\@captype{figure}%
\noindent\begin{minipage}{0.999\linewidth}\begin{center}}
{\end{center}\end{minipage}\smallskip}
\makeatother

\begin{document}

\title{A flux-limited sample of $z\sim1$ Ly$\alpha$ emitting 
galaxies in the CDFS\altaffilmark{1,2}}

\lefthead{Barger, Cowie, \& Wold}

\author{
A.~J.~Barger,$\!$\altaffilmark{3,4,5}
L.~L.~Cowie,$\!$\altaffilmark{5}
I.~G.~B.~Wold $\!$\altaffilmark{3}
}

\altaffiltext{1}{Based in part on data obtained from the Multimission
Archive at the Space Telescope Science Institute (MAST).  STScI is
operated by the Association of Universities for Research in Astronomy,
Inc., under NASA contract NAS5-26555.  Support for MAST for non-HST
data is provided by the NASA Office of Space Science via grant
NAG5-7584 and by other grants and contracts.}
\altaffiltext{2}{This research used the facilities of the Canadian Astronomy 
Data Centre operated by the National Research Council of 
Canada with the support of the Canadian Space Agency.}
\altaffiltext{3}{Department of Astronomy, University of
Wisconsin-Madison, 475 North Charter Street, Madison, WI 53706.}
\altaffiltext{4}{Department of Physics and Astronomy,
University of Hawaii, 2505 Correa Road, Honolulu, HI 96822.}
\altaffiltext{5}{Institute for Astronomy, University of Hawaii,
2680 Woodlawn Drive, Honolulu, HI 96822.}

\slugcomment{Accepted by The Astrophysical Journal}


\begin{abstract}

We describe a method for obtaining a flux-limited sample of \lya\ 
emitters from {\em GALEX\/} grism data. We show that the multiple
{\em GALEX\/} grism images can be converted into a three-dimensional
(two spatial axes and one wavelength axis) data cube. The wavelength
slices may then be treated as narrowband images and searched
for emission-line galaxies. For the {\em GALEX\/} NUV grism data,
the method provides a \lya\ flux-limited
sample over the redshift range $z=0.67-1.16$. We test the method
on the {\em Chandra\/} Deep Field South field, 
where we find 28 \lya\ emitters
with faint continuum magnitudes (NUV$>22$) that are not present
in the {\em GALEX\/} pipeline sample. We measure the completeness
by adding artificial emitters and measuring the fraction recovered.
We find that we have an $80\%$ completeness above a \lya\ flux
of $10^{-15}$~erg~cm$^{-2}$~s$^{-1}$.
We use the UV spectra and the available X-ray data and optical spectra 
to estimate the fraction 
of active galactic nuclei in the selection. We report the
first detection of a giant \lya\ blob at $z<1$, though we find that
these objects are much less common at $z=1$ than at $z=3$.
Finally, we compute limits on the $z\sim1$ \lya\
luminosity function and confirm that there is a dramatic evolution 
in the luminosity function over the redshift range $z=0-1$.

\end{abstract}


\section{Introduction}

Ly$\alpha$\ is the only emission line that we can detect in the
highest redshift ($z>6$) galaxies, making it the only probe of
the internal structure of these galaxies and one of the few
diagnostic tools for studying the intergalactic gas.
However, separating the internal and external
effects will always be difficult at these redshifts.
Thus, to determine the intrinsic Ly$\alpha$\ properties of 
galaxies, we need to study lower redshift samples where 
the galaxy structure can be separated from the effects of the 
intergalactic medium.

Recently, substantial samples of $z\sim0.2-0.4$ \lya\ emitters (LAEs) 
and a small number of $z\sim1$ LAEs have been 
found with the {\em Galaxy Evolution Explorer (GALEX};
Martin et al.\ 2005) grism spectrographs. 
The pipeline extractions for the grism spectrographs 
provide spectra to a fixed limiting near-ultraviolet (NUV) 
magnitude that can be searched for \lya\ emission lines
(Deharveng et al.\ 2008; Cowie et al.\ 2010).
The procedure enables the selection of a substantial sample of 
sources that can be compared to the high-redshift LAEs.  
Since the selection process is based on finding \lya\ in the 
UV-continuum selected {\em GALEX\/} sources, it is most analogous 
to locating LAEs in the Lyman break galaxy (LBG) 
population using spectroscopy (e.g., Shapley et al.\ 2003).

The advent of these low-redshift LAE catalogs has been very 
exciting, and many follow-up papers have been written
(e.g., Finkelstein et al.\ 2009a, 2009b; 
Atek et al.\ 2009; Scarlata et al.\ 2009; Cowie et al.\ 2011).
However, because the {\em GALEX\/} grism pipeline only includes
objects whose UV continuum magnitudes are bright enough to generate 
measurable continuum spectra, only  a fraction of the emission-line
objects at these redshifts are detected.  
In particular, strong emitters with high 
equivalent widths (EWs) may have emission lines that are detectable but 
continuum magnitudes that fall below the continuum threshold. 

Moreover, because the samples have both a continuum and an emission-line 
selection, it is not straightforward to compute the LAE luminosity
functions (see the extensive discussions in Deharveng et al.\ 2008 and 
Cowie et al.\ 2010). Thus, ad hoc assumptions about the invariance of the
EW distribution as a function of continuum magnitude
need to be made, even though they may not be correct.

Fortunately, it is possible to detect directly the excluded emitters
by re-extracting the {\em GALEX\/} data.  In this paper, we show
how to find ``orphan'' emission lines 
(e.g., Straughn et al.\ 2008; Keel et al.\ 2009)
and how to generate a flux-limited \lya\ emission-line sample.
We refer to this approach as a data cube search.
First, we extract a spectrum for every spatial pixel in the {\em GALEX\/} 
fields. Next, we form a three-dimensional data cube, analogous
to that produced by an integral field unit, with two spatial axes 
and a third wavelength axis.  Finally, we search the wavelength slices 
in these data cubes for emitters. 

An alternative procedure is to use the deep broadband {\em GALEX\/} 
images to develop a target list of galaxies fainter than those used
in the pipeline and then to extract the spectra of these objects. 
However, since this approach still introduces a 
pre-selection (which must be included and modeled in any interpretation
and may omit extreme EW emitters), we prefer to use the data 
cube approach.

In order to fully understand the LAEs that we find, we need a 
wealth of ancillary data. In particular, the key requirement is as 
extensive a set of redshift measurements as possible.  
Thus, for this paper, we focus on the intensively studied
{\em Chandra\/} Deep Field South (CDFS; Giacconi et al.\ 2002) region, 
where we can use both existing optical spectroscopy to confirm the UV 
identifications of the \lya\ line and deep X-ray data
to identify active galactic nuclei (AGNs). 
In Section~\ref{choice}, we describe the rich ancillary 
data on the CDFS. In Section~\ref{technical}, we detail our
data cube search methodology and use it to extract a 
sample of LAEs at $z\sim1$. We also use simulations to estimate
the completeness with which we can detect emitters
of a given flux.
In Section~\ref{newLAE}, we present the properties of the
28 new $z\sim1$ LAEs that we found in our data cube search.
We make use of the ancillary data to test the reliability of the LAEs 
and their UV redshifts and to get a first estimate of the AGN contamination.  
In Section~\ref{disc}, we present our discovery of a giant \lya\
blob. We also compute the $z\sim1$ LAE luminosity function, which
we compare with LAE luminosity functions at lower and higher redshifts.  
In Section~\ref{summary}, we summarize our results.  

We use a standard $H_0=70$~km~s$^{-1}$~Mpc$^{-1}$, 
$\Omega=0.3$, $\Omega=0.7$ cosmology throughout. All magnitudes are 
in the AB magnitude system.

\section{Choice of Field}
\label{choice}

The four deepest NUV grism observations 
(CDFS-00, 353~ks; GROTH-00, 291~ks; NGPDWS-00, 165~ks; 
COSMOS-00, 140~ks) 
correspond to some of the most intensively studied 
regions in the sky.  In this paper, we use the central
$50'$ by $50'$ region (centered on right ascension 
$3^{\rm h} 30^{\rm m} 40.2^{\rm s}$ and declination $-27^\circ 27' 43.4''$ 
in J2000 coordinates) of the CDFS-00 for a pilot study to show
the power of the data cube method.
This region is extremely rich in ancillary information.
For example, it contains the {\em HST\/} ACS Great Observatories 
Origins Deep Survey-South 
(GOODS-S; Giavalisco et al.\ 2004) subregion, where there are more 
than 3000 spectroscopic redshifts in a 140~arcmin$^2$ area
(Le F{\`e}vre et al.\ 2005; Mignoli et al.\ 2005; Vanzella et al.\ 2008; 
Popesso et al.\ 2009; Balestra et al.\ 2010; Cooper et al.\ 2012;
Cowie et al.\ 2012).
In Figure~\ref{spectra}, we show how $>80\%$ of the galaxies with 
$B<24$ in the GOODS-S have spectroscopic identifications. 
In addition, multicolor observations from the {\em HST\/} programs
GOODS-S (Giavalisco et al.\ 2004), 
Cosmic Assembly Near-Infrared Deep Extragalactic Legacy Survey
(CANDELS; Grogin et al.\ 2011; Koekemoer et al.\ 2011), 
and Galaxy Evolution from Morphologies and SEDs
(GEMS; Rix et al.\ 2004) provide the galaxy morphologies and 
spectral energy distributions from the rest-frame far-UV (FUV) to the 
mid-infrared for a substantial fraction of the objects. 
Finally, the 4~Ms {\em Chandra\/} image of the CDFS 
(Luo et al.\ 2008; Xue et al.\ 2011) 
and the Extended CDFS (ECDFS; Lehmer et al.\ 2005; Virani et al.\ 2006) 
fields can be used to help eliminate AGNs. A number of optical
spectroscopic surveys have specifically targeted the X-ray sources 
in these fields (Szokoly et al.\ 2004; Treister et al.\ 2009; 
Silverman et al.\ 2010).

\begin{inlinefigure}
\centerline{\includegraphics[width=3.8in]{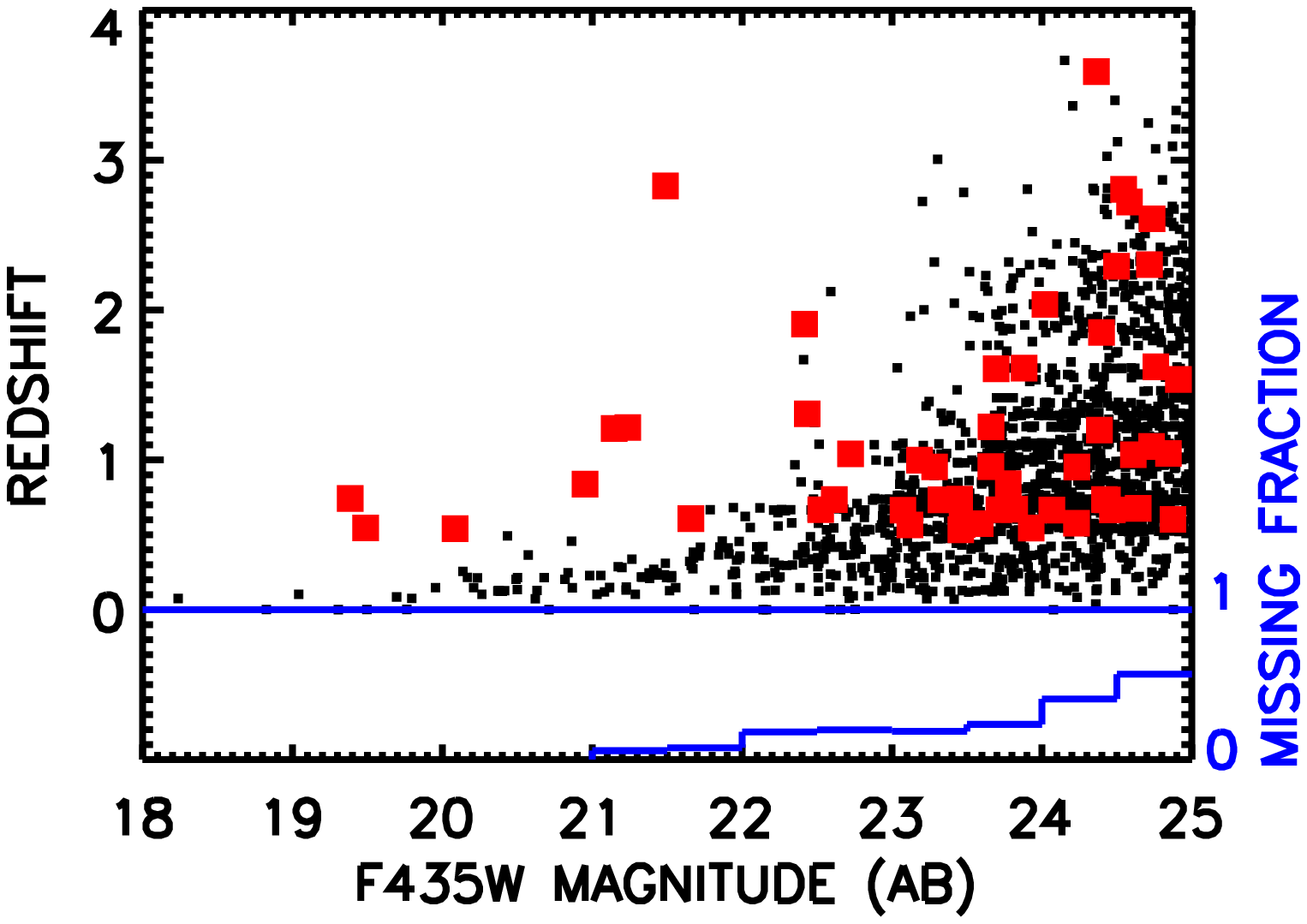}}
\caption{
Spectroscopic redshift vs. F435W ($B$-band)
magnitude for the sources in the GOODS-S (see references in text). 
The bottom blue histogram shows the fraction of sources without 
secure spectroscopic identifications. This fraction only becomes
large fainter than $B=24$. Red squares 
show sources classified as AGNs based on their X-ray 
luminosities (see Section~\ref{newLAE}).
\label{spectra}
}
\end{inlinefigure}

\section{Methodology}
\label{technical}

\subsection{GALEX Spectral Images}
\label{images}

The 50~cm {\em GALEX\/} telescope contains a selectable CaF2 
75~g~mm$^{-1}$ grism, which provides slitless spectroscopy over the 
whole {\em GALEX\/} 1.25~deg diameter field-of-view.  The grism is 
optimized for order~2 in the FUV channel ($1344-1786$~\AA) and for 
order~1 in the NUV channel ($1771-2831$~\AA).
The resolution is roughly 10~\AA\ in the FUV (R=200) and 25~\AA\ in the 
NUV (R=90), though the exact resolution depends on both the wavelength 
and the spatial extent of the target (Morrissey et al.\ 2007).

The pipeline extractions of the {\em GALEX\/} grism data are described in 
Chapter~6 of the {\em GALEX\/} technical documentation and in the 
{\em GALEX\/} grism primer
(\url{http://www.galex.caltech.edu/researcher/grism/primer.html}).
The pipeline produces spectral images for each individual exposure,
together with extracted one- and two-dimensional spectra from the
combined exposures for all of the objects in the field down to 
a limiting NUV magnitude where the signal-to-noise of the continuum 
spectrum becomes low. 

In the first stage of the {\em GALEX\/} grism reductions, the photons 
are processed in the same way as they are in the direct broadband images,
and astrometrically corrected and registered whole field spectral images 
of the intensity and system response are created. 
These individual spectral images, with an average exposure time
of about a thousand seconds, have been taken at a wide range of grism 
rotations.  For the long total exposures of fields like the CDFS-00, 
there are several hundred such individual spectral images.  We begin 
our own data reductions from these spectral images.

\subsection{Data Cube Construction}
\label{constructdatacubes}

The slitless nature of the grism data results in spectral overlaps
between neighboring objects.  However, because each 
of the spectral images is taken at a different rotation angle, overlaps
can be removed.  Objects only overlap with a given neighbor
at one orientation, and so in the multiple spectral images, the signal
associated with a given object can be picked out from the occasional
contamination.  The grism data are often portrayed as a 
``propellor image'', where images at many grism angles are added 
together, as shown in Figure~\ref{propellor}.  
This type of image is extremely useful for visualizing our present 
{\em GALEX\/} grism extraction process.  Each pixel (x, y)
on the spectral image corresponds to a specified position in right
ascension and declination.
At a given orientation, the grism spectrum corresponding to that 
pixel runs at that angle relative
to the central position. The zeroth order lies to the negative
side, while the higher orders lie to the positive side. 
Thus, for intensively observed fields with large numbers of 
independent spectral images, we can form a four-dimensional data
cube (x\,\,by\,\,y\,\,by\,\,the number of wavelength elements\,\,by\,the
number of spectral images). 
The number of x and y pixels corresponds to the
total number of spatial pixels in the grism image (here 2000 by 2000),
and we hereafter refer to the number of spectral images as NN.

\begin{inlinefigure}
\vskip 0.4cm
\centerline{\includegraphics[width=2.8in,angle=90]{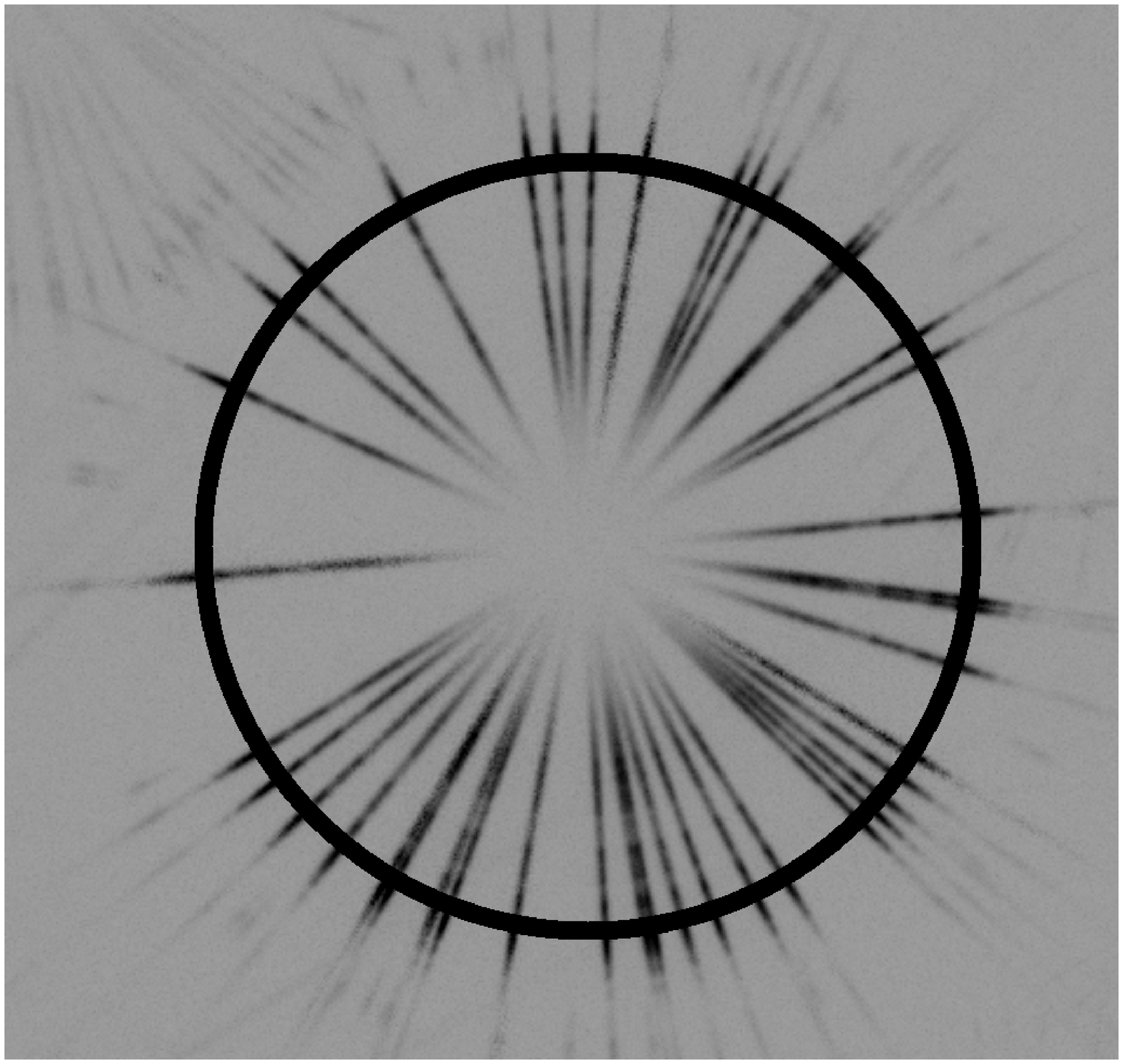}}
\caption{
The well-known ``propellor'' image for a small
subsample of spectral images in the CDFS-00. 
The image is the sum of a number of individual spectral images.
We show only 40 of the 342 spectral images on the field. Each 
exposure is taken at a different angle. In a summed
image such as this, they fan out in a propellor shape centered
on the position of the galaxy (which is not seen in the
grism spectra). For the NUV grism images shown here, the
spectrum is dominated by the first order.  For this order,
a given wavelength corresponds to a
circular annulus, such as that shown by the large circle.
The flux at that wavelength may be obtained
by forming a clipped average of the vector of
fluxes in the region intercepted by the annulus and the spectra
for the full image set (e.g., Keel et al.\ 2009).
\label{propellor}
}
\end{inlinefigure}

A given wavelength corresponds to a circular annulus (large circle
in Figure~\ref{propellor}), the spokes
of which correspond to the grism angles. A given emission
line will occur at these specified positions in the suite of images 
with varying grism angles (e.g., Keel et al.\ 2009).
If we construct a vector of values, only
the object itself appears in all the elements. Other objects
occur as random contamination in single elements of the vector.
Thus, we can mask out pixels on the individual spectral images 
where we expect contamination from bright neighboring galaxies. 
We can further remove contamination from fainter objects
by using a clipped average to determine the true signal.
In doing so, we can collapse the four-dimensional data cube
to a three-dimensional data cube (x\,\,by\,\,y\,\,by\,\,the number 
of wavelength elements). This final three-dimensional
data cube is analogous to the
data cube that would be obtained with a scanning imaging spectrometer
or with an integral-field spectrograph where each spatial pixel
has a spectrum associated with it. For most spatial pixels, there is 
no significant continuum flux or emission lines in the spectrum
(the third axis). We may also view the three-dimensional
data cube as a sequence of narrowband images, stepping in
wavelength, each of which may be searched for emitters. It is in 
this way that we form our emission-line selected sample.

We created a pipeline to convert the NUV and FUV spectral images into 
four-dimensional data cubes. We chose a wavelength step of 5~\AA\ for 
the NUV output cube in order to oversample the roughly 20~\AA\ 
wavelength resolution. We chose the wavelength range ($1930-2730$~\AA)
to cover the 25\% response range of the NUV grism.
This gives 160 wavelength elements. 

In forming the four-dimensional data cubes 
(x\,\,by\,\,y\,\,by\,\,160\,\,by\,NN), we wanted to maintain 
the statistical independence of the pixels. Thus,
in assigning to each (x, y, $\lambda$) a vector 
(the fourth axis), 
we need to be concerned with the fact that we are oversampling
the spectral images in wavelength. 
To avoid averaging, we decided to assign
to each (x, y, $\lambda$) position in a given spectral image
(recall that each position will have NN values since there are 
NN spectral images) the nearest neighbor pixel
using the grism angle from the {\em GALEX\/} pipeline.
Note that the {\em GALEX\/} pipeline computes the grism angle by 
fitting to the spectra of bright objects in the field.
To maintain the statistical independence of the pixels, 
we will only assign each 
pixel in a given spectral image once, so since we
are oversampling the spectral images in wavelength, this means 
that some of the pixels in the fourth axis are blank. 
These blank pixels are ignored hereafter.

At this point, we collapsed the fourth axis by forming the median,
comparing each value in the fourth axis with the median, and
eliminating points that deviate from the median by more than 
$5\sigma$. We formed an exposure time-weighted average from the 
remaining points to give the final signal at this (x, y, $\lambda$) 
pixel in the three-dimensional data cube. 

We used an identical procedure for the FUV processing, except here 
we chose a wavelength step of 2.5~\AA\ over the wavelength range 
$1345-1745$~\AA, giving 160 wavelength elements. Also, because
of the low background photon detection rate per pixel in the FUV,
and because contamination is much less of a problem in the FUV,
we formed a simple average rather than a median for comparison
with each value in the fourth axis. We eliminated points that 
deviated from the average by more than $5\sigma$.
Then, as we did for the NUV processing, we formed
an exposure time-weighted average of the clipped vector 
to obtain the final signal at this (x, y, $\lambda$) pixel in the 
three-dimensional data cube.

For the present emission-line search, we are not interested
in the objects with brighter continua whose spectra are already
included in the existing {\em GALEX\/} pipeline data products. 
We also want to  minimize any contamination from these brighter
galaxies and stars in the final three-dimensional data cube. 
We therefore masked out the spectral region in each
{\em GALEX\/} spectral image that corresponded to a pixel with a 
surface brightness in excess of 25.8~mag~arcsec$^{-2}$
in the NUV continuum image of the field smoothed with a 
2 pixel boxcar smoothing. 
This removes all objects with NUV magnitudes brighter than 20.4 
and no objects with NUV magnitudes fainter than 21.25. This faint 
magnitude bound is considerably brighter than the magnitude limit 
of the pipeline sample, so we only eliminate objects that already 
have spectra and hence already have been searched
for \lya\ lines. (The remaining objects that are already present
in the pipeline sample are eliminated later in the process; see
Section~\ref{searchdatacubes}.)
None of our results are sensitive to
the precise choice of the masking surface brightness threshold.

In Figure~\ref{slice_sample} (left panel), we show a typical 
portion of a wavelength ``slice'' (a 40~\AA\ region obtained by
summing eight of the 5~\AA\ steps; see Section~\ref{searchdatacubes}) 
in the final data cube 
compared to (right panel) the corresponding NUV continuum image.
The positions in the final data cube that correspond to the 
regions that were masked appear white in the wavelength
slice image. 10901 spatial pixels are eliminated by the masking,
or just under $0.3\%$ of the spatial area covered by the 
three-dimensional data cube.

\subsection{Data Cube Search}
\label{searchdatacubes}

\begin{figure*}
\centerline{\includegraphics[width=2.8in,angle=90]{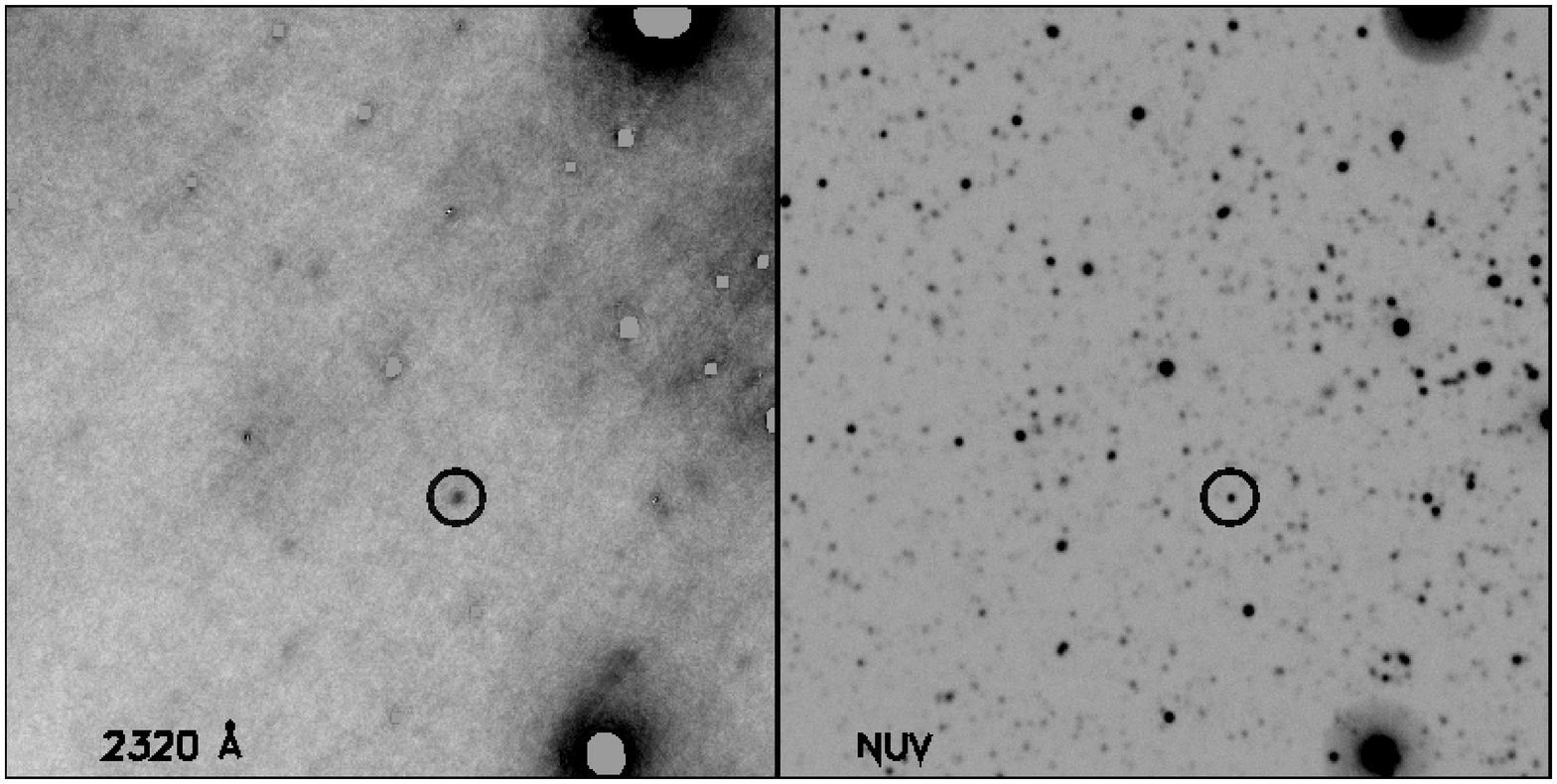}}
\caption{{\em (Left)\/} A 40~\AA\ image slice (centered at 2320~\AA) 
of the CDFS-00 NUV data cube.
The spatial area is $10'$ by $10'$, centered at a right ascension of 
$3^{\rm h} 32^{\rm m} 10.18^{\rm s}$ and a declination of 
$-28^\circ 12' 46\farcs1$.  Brighter galaxies in 
the field have been zeroed out in this data cube extraction, as described 
in the text, though there are still residuals around the edges 
of these masked regions that can cause contamination. 
We mask these regions at a later stage. The variation in the background
is caused by the relative positions of bright stars in the field.
The circled object is an emitter, GALEX033202-281408, which we 
use to illustrate our search procedure in Figure~\ref{image_slice}.
{\em (Right)\/} Corresponding NUV continuum image with the
emitter circled.
\label{slice_sample}
}
\end{figure*}

\begin{figure*}
\centerline{\includegraphics[width=3.7in,angle=90]{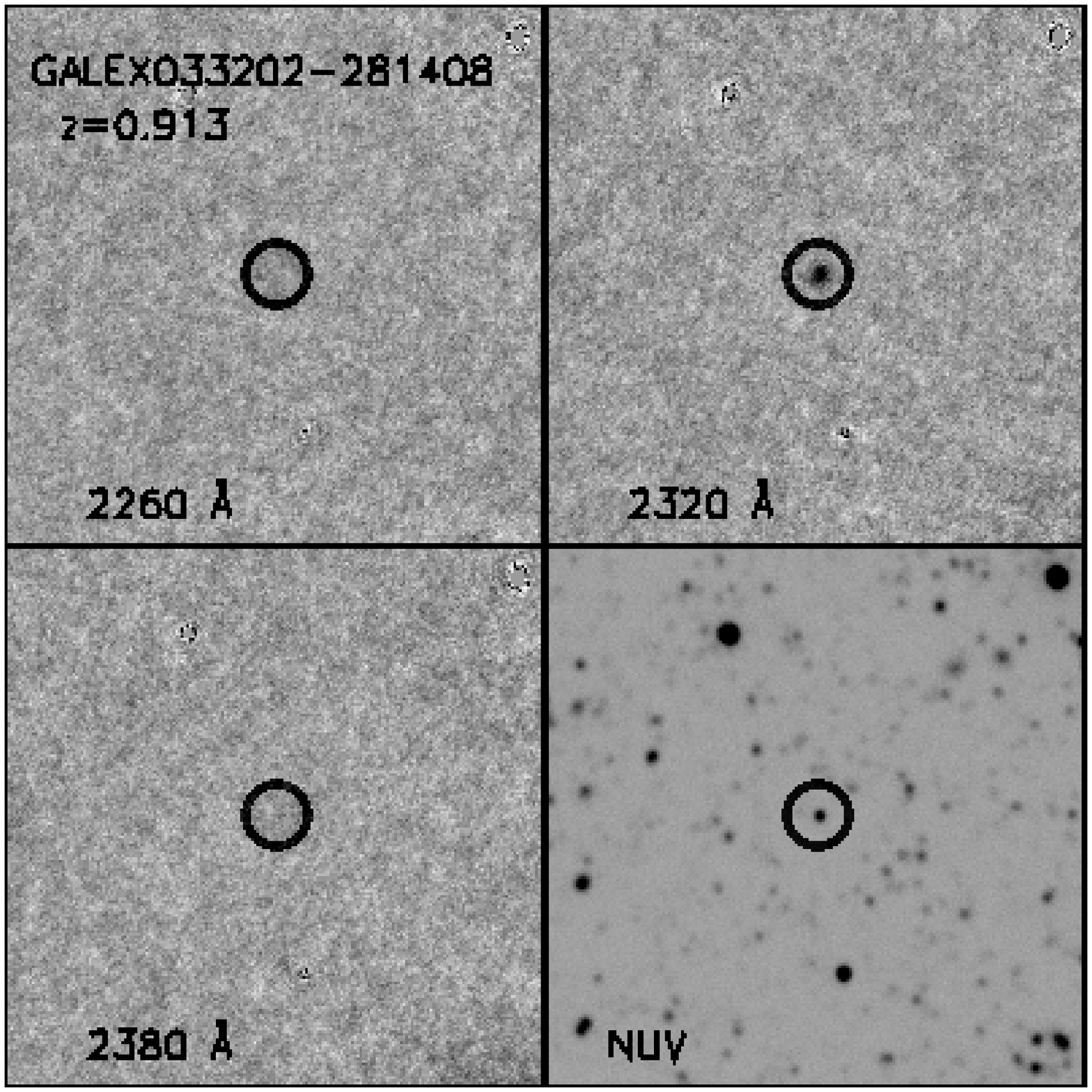}}
\caption{
Three $5'\times 5'$ background subtracted 20~\AA\
image slices and the corresponding NUV continuum image
to illustrate the search procedure for finding LAEs in the 
CDFS-00 NUV data cube.  The images
correspond to a portion of the region shown in Figure~\ref{slice_sample}
and are centered on the detected LAE GALEX033202-281408, which
is circled. This LAE has
a flux of $3.1\times10^{-15}$~erg~cm$^{-2}$~s$^{-1}$ 
and a UV spectral redshift of $z=0.913$. It appears 
in the image slice at 2320~\AA\ but not in the off emission-line 
slices at 2260~\AA\ and 2380~\AA. It corresponds
to a relatively bright object (21.9~mag) in the NUV continuum image
and only just misses being included in the {\em GALEX\/} pipeline sample.
\label{image_slice}
}
\end{figure*}

With the three-dimensional data cube in hand, we can now carry 
out a search for emission-line objects. We first divided 
each 5~\AA\ wavelength slice in the data cube by the grism 
effective area at that wavelength to take out response variations.
We used the values given in 
\url{http://galexgi.gsfc.nasa.gov/docs/galex/Documents}. 
We then formed forty 20~\AA\ (in order to roughly match the 
resolution element) narrowband slices from the NUV data cube.
For each slice, we subtracted the average
of slices on either side of the primary slice, N. We used slices
N$-5$ to N$-3$ and slices N$+3$ to N$+5$ to form this average. 
The background subtraction removes most of the residual structure 
and most of the continuum objects in the field, as we see
by comparing the left panel of Figure~\ref{slice_sample} with
the upper-right panel of Figure~\ref{image_slice}. 
This will retain any emitter whose width is less than about 240~\AA.  
Even broad-line AGNs satisfy this criterion, so we should not
be eliminating any objects by this process.

From the left panel of Figure~\ref{slice_sample}, we can see 
residual regions around the small number of very bright 
sources that were masked, and these can cause contamination.  
Thus, in each of the 40 slices, we masked out additional spatial 
regions around these bright sources.  
For sources with NUV magnitudes brighter than 14, 
we masked a $4'$ radius region around the source, and for sources 
with NUV magnitudes brighter than 17, we masked a $1.5'$ radius region.

We searched each slice for objects
with surface brightnesses of at least 
$2\times10^{-6}$~photons~cm$^{-2}$~s$^{-1}$ per 
$1.5''$ spatial pixel using a 4 pixel spatially smoothed version 
of the slice. The smoothing roughly matches the $5.3''$
spatial resolution in the {\em GALEX\/} NUV band.  
The surface brightness cut was chosen empirically to provide
the maximum depth while not generating an unmanageably
large number of spurious sources that have to be
eliminated by hand.
Bright emitters may appear in more than one slice, and
we formed our preliminary catalog by merging
all of the emitters found in neighboring slices. 
As a consequence of the background subtraction process, we
only cover a wavelength range of 2030 to 2630~\AA, which
corresponds to a redshift range for \lya\ of $z=0.67-1.16$. 
At this stage, any objects already in the pipeline sample were 
removed. 

%

We created the one-dimensional spectra directly from the three-dimensional 
data cube using a $4\times4$ spatial pixel box together with a local 
background subtraction. We flux calibrated by matching to the NUV 
magnitude of the object.  We visually inspected each of the narrowband 
slices and the one-dimensional spectrum of each selected object and 
eliminated objects that were artifacts, such as remaining edge effects 
from the brighter objects  or objects with stellar spectra,
to form our primary catalog. The initial selection yielded 161 objects, 
of which we retained 28. 

For each of the 28 objects, we also created a two-dimensional spectral 
image from which we could extract final one-dimensional spectra. 
For doing the extractions, we created our own version of the {\em GALEX\/} 
pipeline.  The only difference between our version of the pipeline and 
the standard pipeline is that we used a profile-weighted spectral 
extraction to form the one-dimensional spectra
(Horne 1986). As we illustrate in Figure~\ref{galex}, we obtained some 
slight improvements in signal-to-noise from the profile weighting (about 
a factor of 1.3), which is helpful in dealing with weak features, such as 
the faint \lya\ emission lines. (Note that the source shown in 
Figure~\ref{galex} was from the pipeline sample rather than from our 
data cube sample.)

\begin{inlinefigure}
\vskip 0.4cm
\centerline{\includegraphics[width=3.8in]{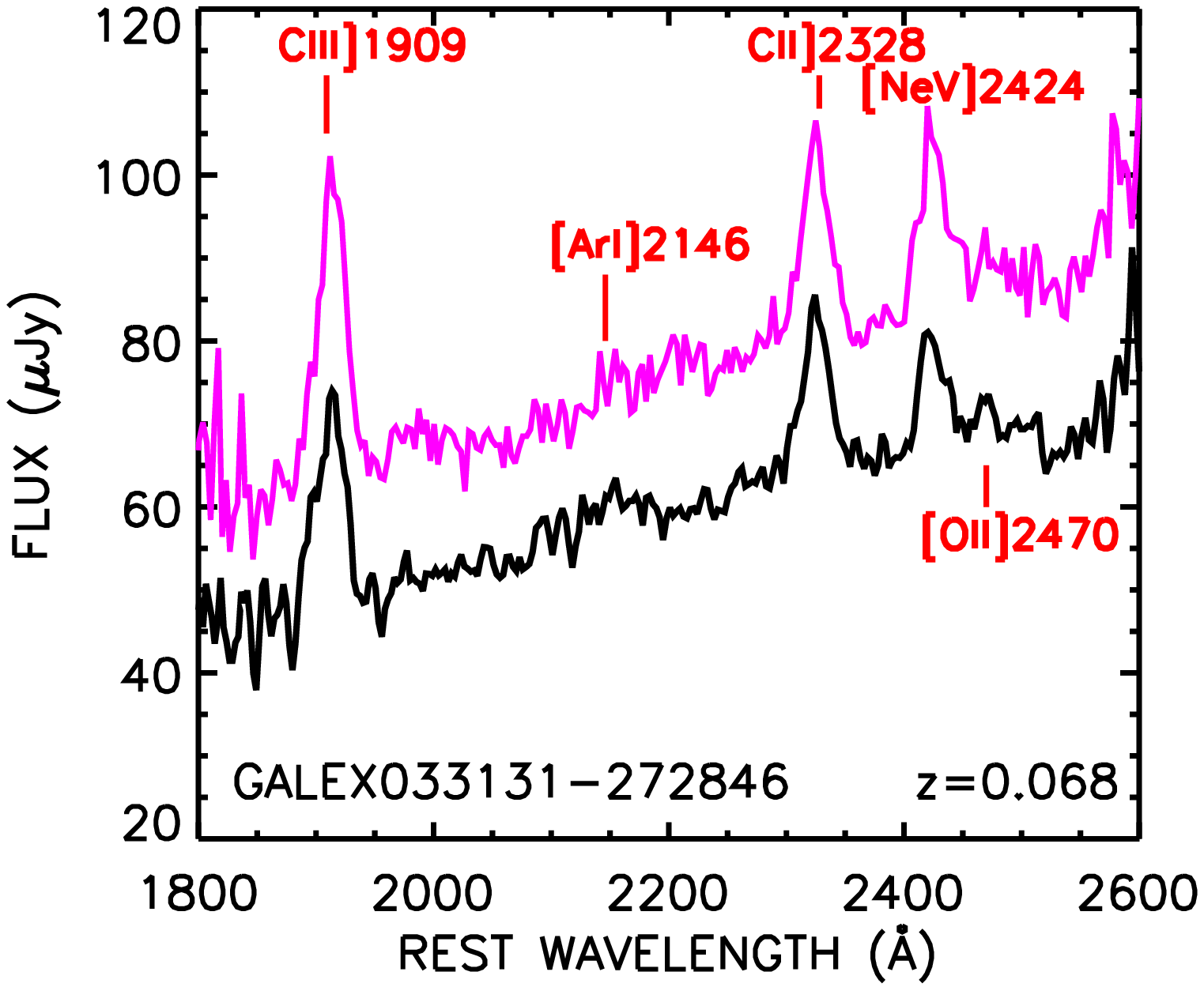}}
\caption{
Spectrum of a low-redshift narrow-line Seyfert found
in the CDFS-00 field. We show only the 
portion of the spectrum covered by the NUV spectrum. The purple
spectrum shows the {\em GALEX\/} pipeline extraction, 
while the black spectrum shows our version, which does a 
profile-weighted extraction (displaced downwards in normalization 
for comparison purposes). The optimal weighting in our extraction 
provides a slight gain (about 30\%) in signal-to-noise,
which allows us to see the weak [OII]$\lambda2470$ 
line. The [ArI]$\lambda2146$ line is still too weak to be seen.
\label{galex}
}
\end{inlinefigure}

\begin{figure*}
\centerline{\includegraphics[width=3.5in,angle=90]{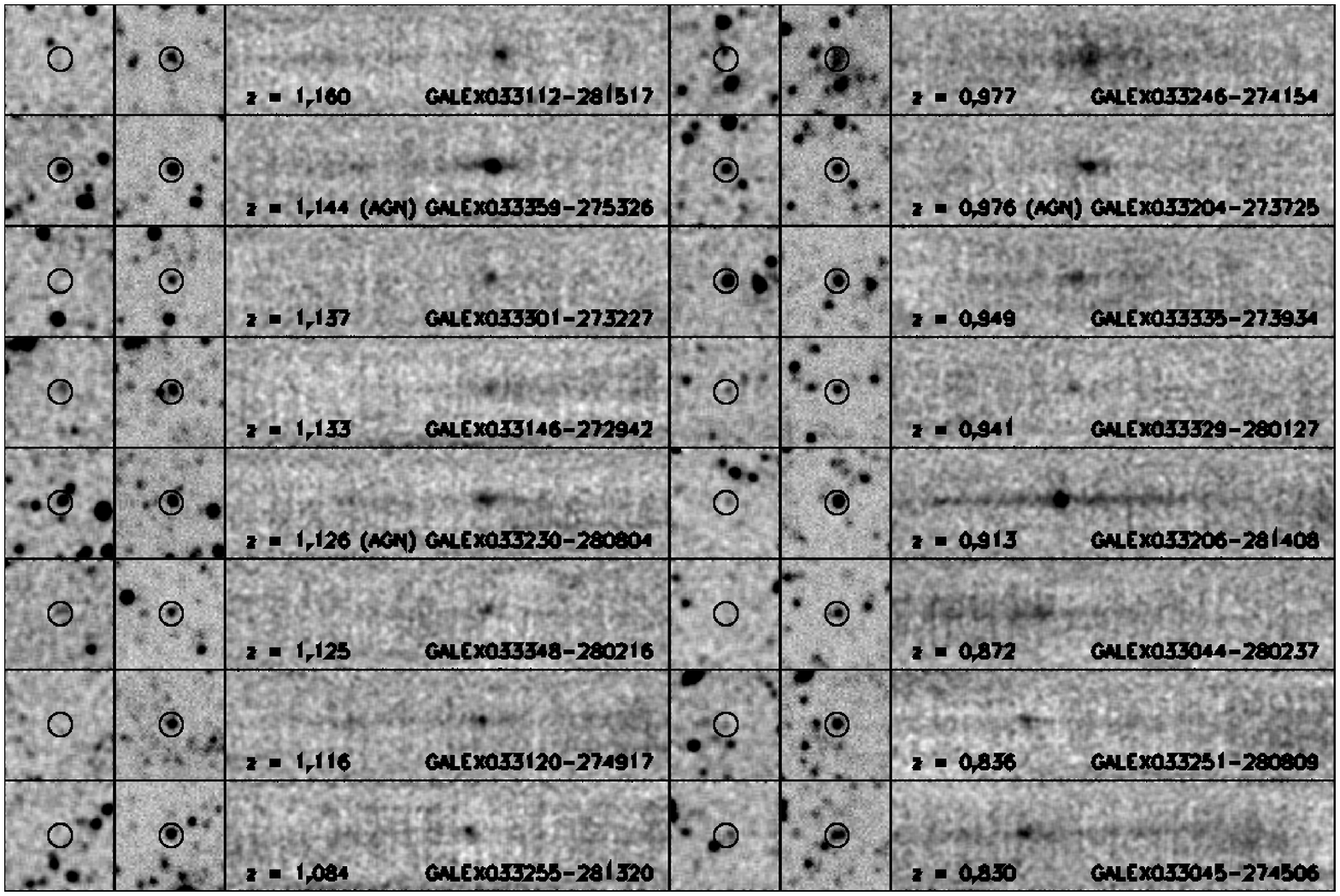}}
\centerline{\includegraphics[width=3.5in,angle=90]{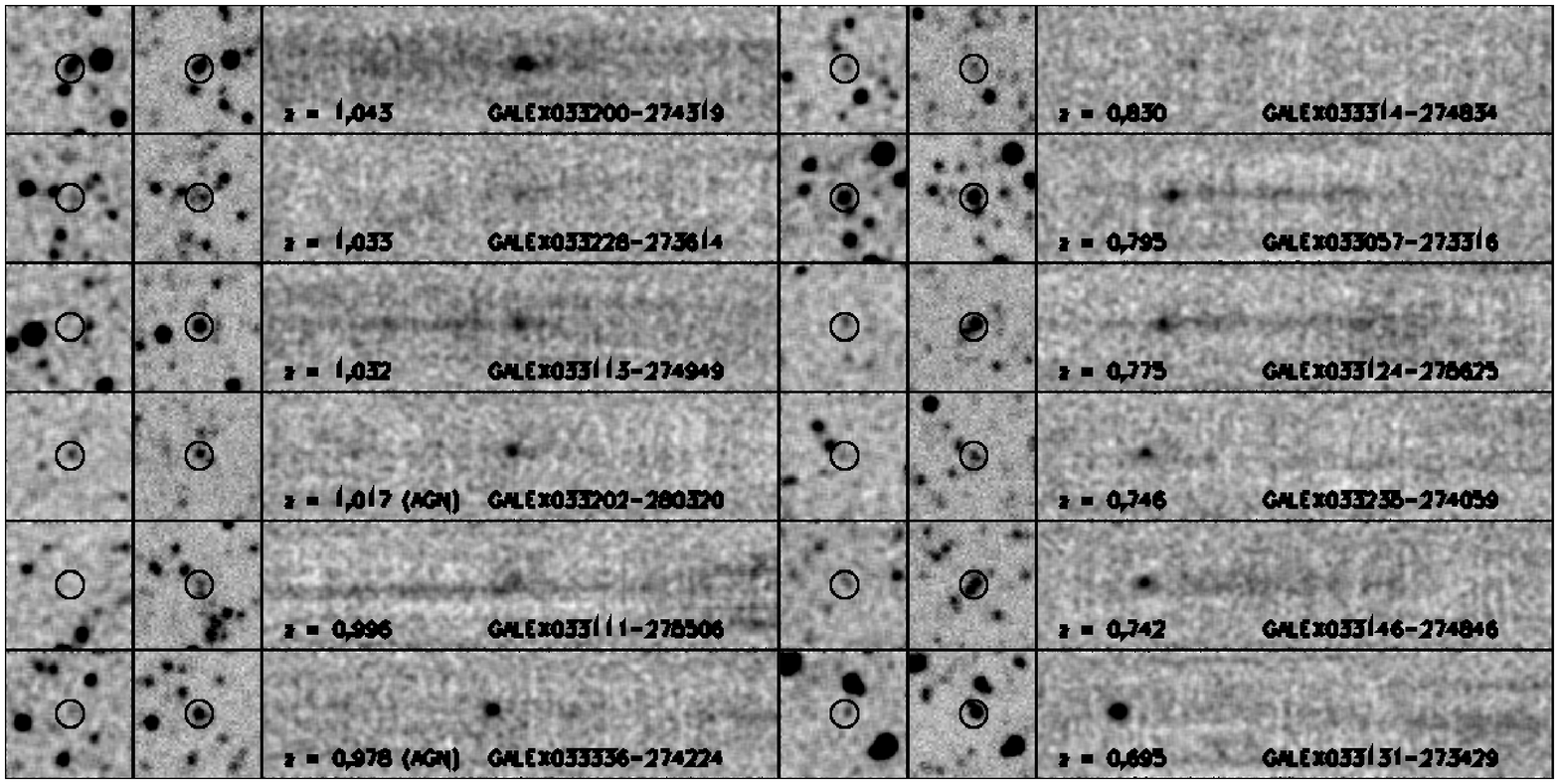}}
\caption{
Two-dimensional spectral images of the 28 new $z\sim1$ LAEs 
found in our CDFS-00 NUV data cube search, ordered by
decreasing redshift, as in Table~\ref{faint_z1_sample_table}. 
The x-axis corresponds to the wavelength range $1892-3119$~\AA, 
and the y-axis corresponds to the spatial dimension.
The spectral images are labeled with the UV spectral redshift;
with ``AGN'', if the source were classified as an AGN based
on the presence of high-excitation lines in the UV spectrum;
and with the {\em GALEX\/} name.
The two small panels at the far-left and left of each spectral image
are broadband FUV and NUV images, respectively, with the position of
the emitter circled in each.
The \lya\ blob is the upper-right corner spectrum.
\label{spectralimages}
}
\end{figure*}

\begin{figure*}
\includegraphics[width=6.5in,angle=0]{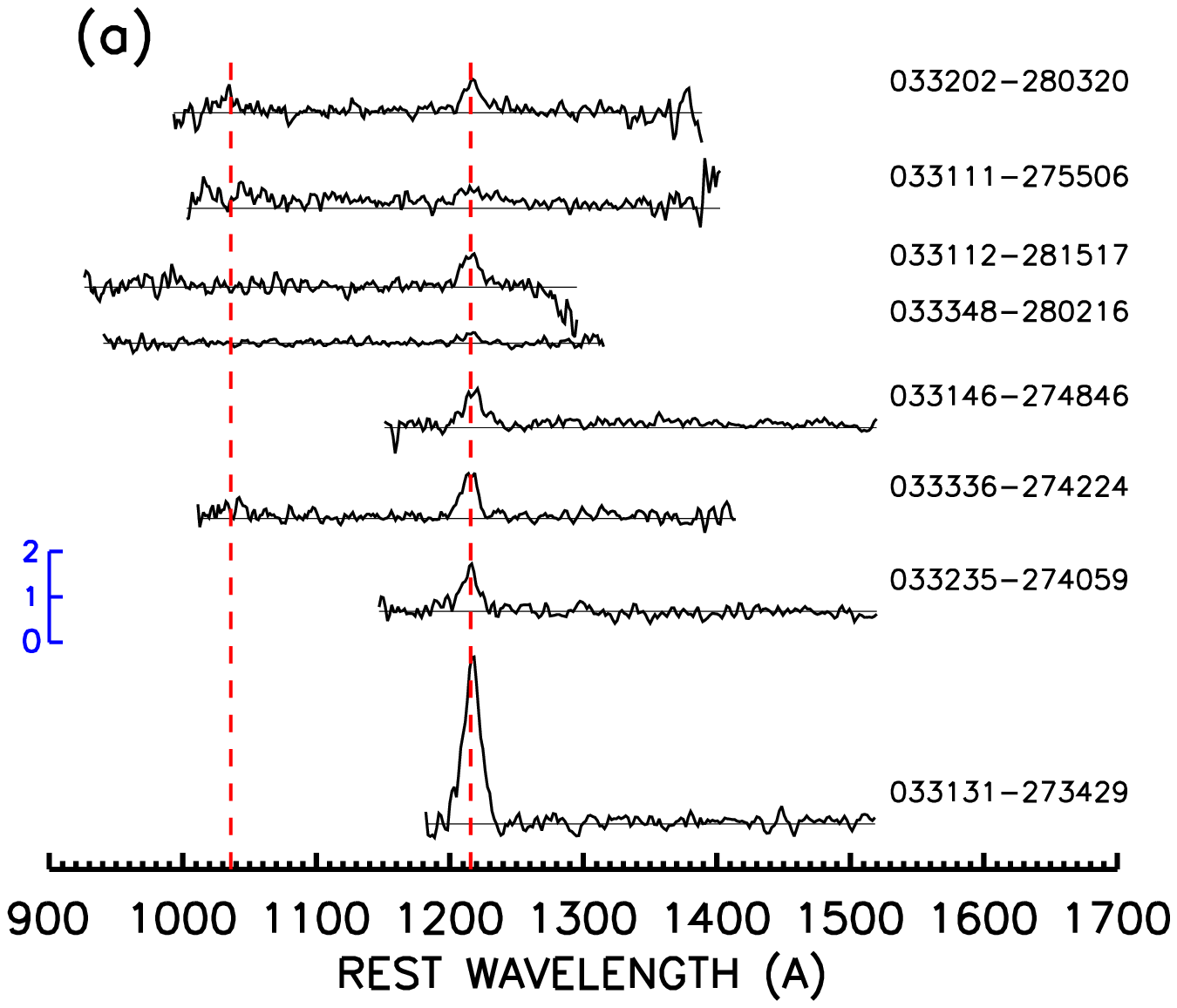}
\caption{(a-d)
One-dimensional spectra of the 28 new $z\sim1$ LAEs found 
in our CDFS-00 NUV data cube search, ordered by decreasing 
redshift, as in Table~\ref{faint_z1_sample_table}. 
The spectra are plotted against rest wavelength and are in units 
of $10^{-16}$~erg~cm$^{-2}$~s$^{-1}$~\AA$^{-1}$,
with the scale shown at the left side of each panel. 
The red dashed lines show the positions of \lya\ and
OVI 1036.
\label{spect}
}
\end{figure*}

\setcounter{figure}{6}
\begin{figure*}
\includegraphics[width=7in,angle=0]{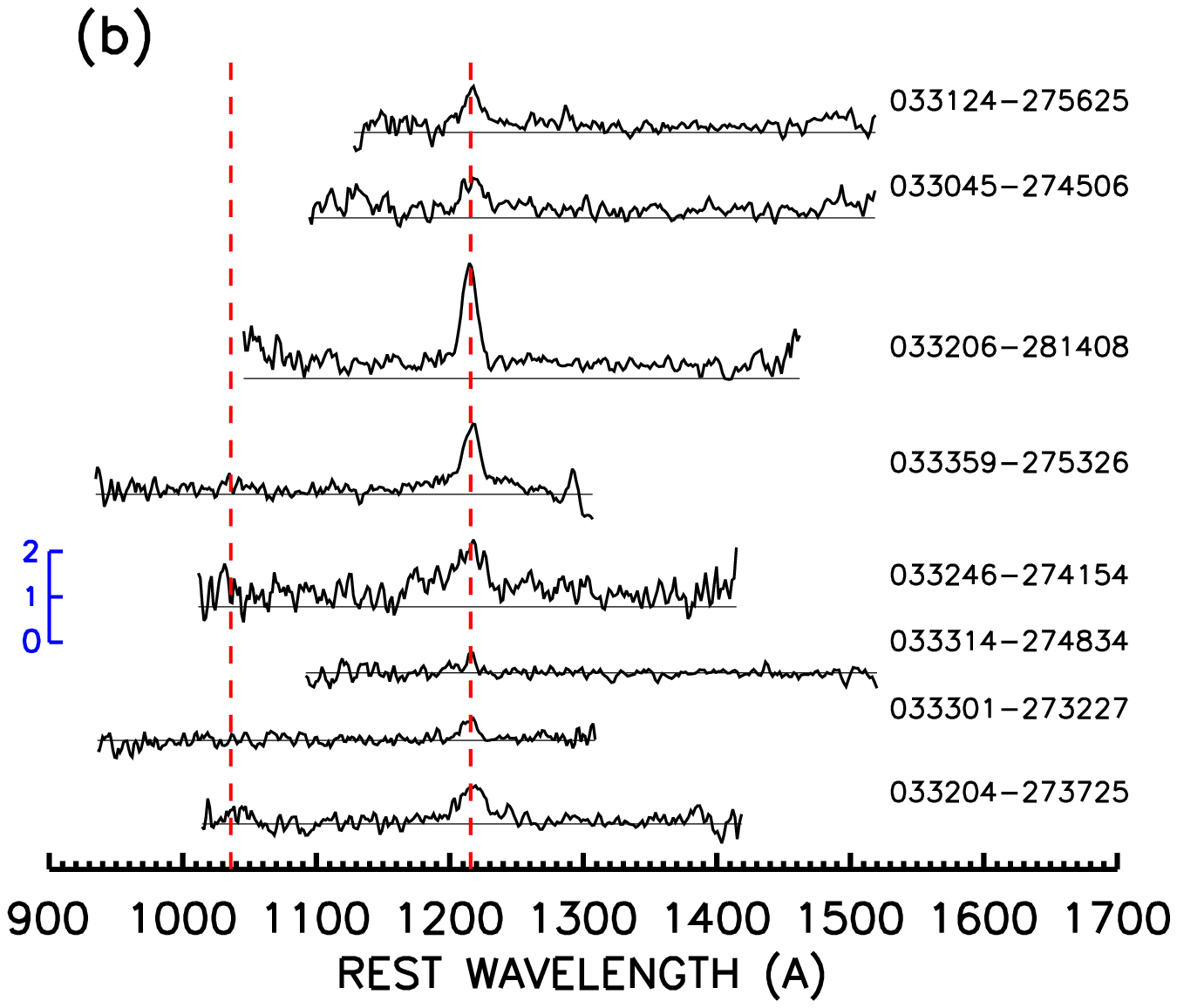}
\caption{
(b)  
\label{spectb}
}
\end{figure*}

\setcounter{figure}{6}
\begin{figure*}
\includegraphics[width=7in,angle=0]{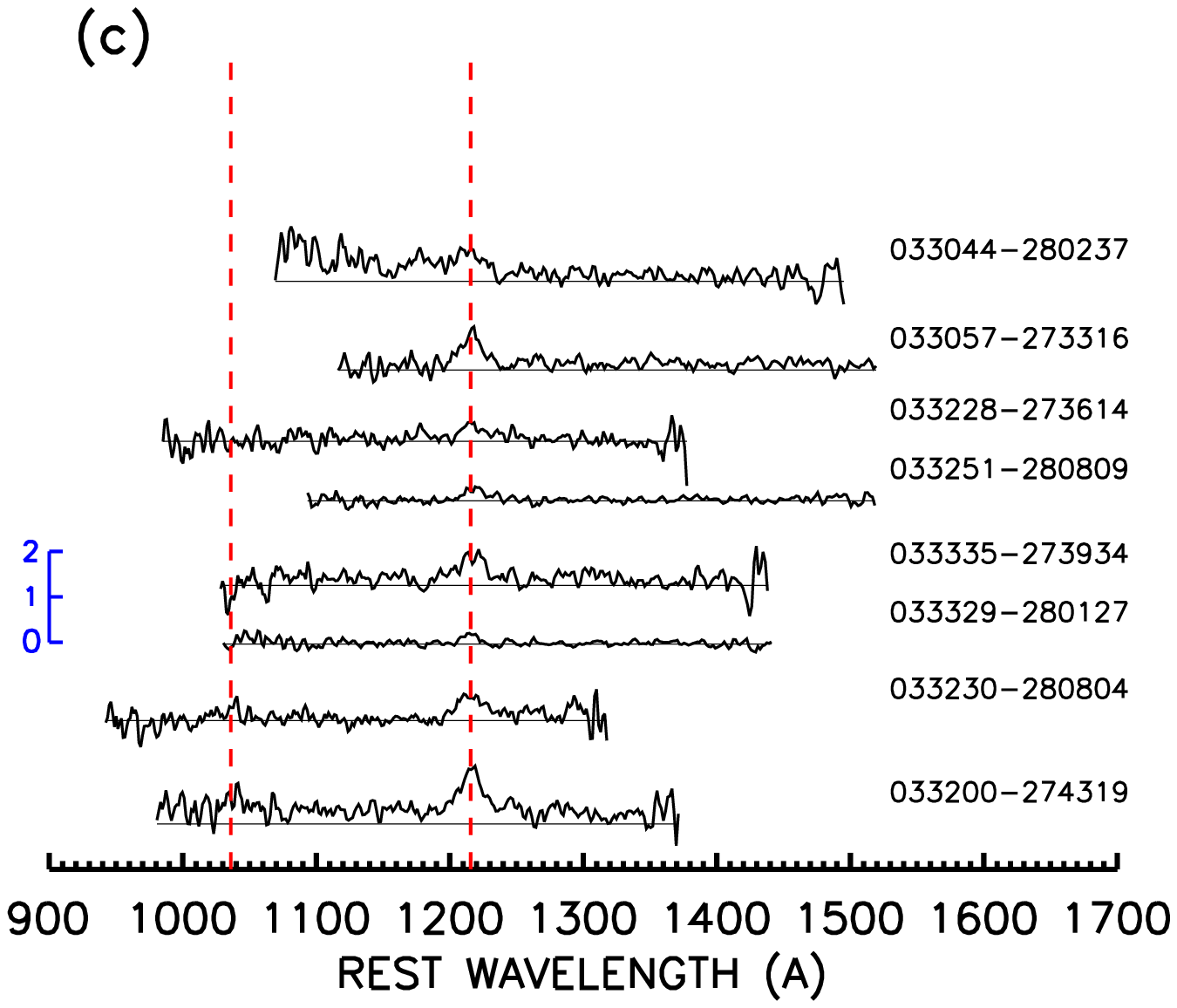}
\caption{
(c)  
\label{spectc}
}
\end{figure*}

\setcounter{figure}{6}
\begin{figure*}
\includegraphics[width=7in,angle=0]{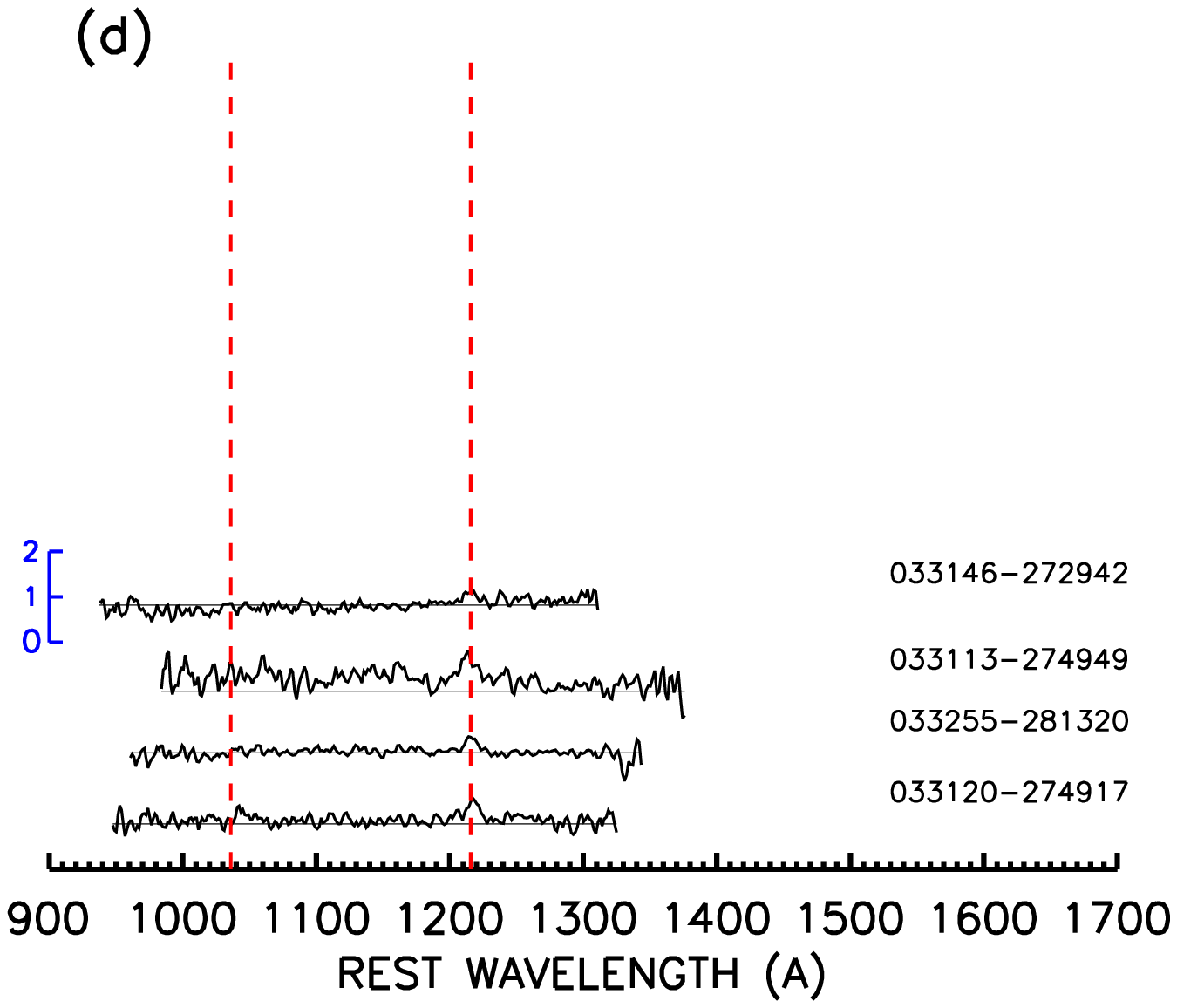}
\caption{
(d)  
\label{spectd}
}
\end{figure*}

In Figure~\ref{spectralimages}, we show the two-dimensional NUV spectral 
images of our 28 new $z\sim 1$ LAEs, ordered according to 
Table~\ref{faint_z1_sample_table} (see Section~\ref{newLAE}),
along with their FUV and NUV continuum images (shown to the far-left 
and left of each spectral image, respectively). 
In Figure~\ref{spect}, we show the one-dimensional spectra that we 
extracted from these.  We performed the final 
flux calibration using the given {\em GALEX\/} spectral response.

We decided that using the one-dimensional spectra formed from our
version of the pipeline software rather than the one-dimensional
spectra obtained from the data cube would enable us to make the most 
direct comparison with brighter emitters in the field, since their 
spectra had been formed from the pipeline software. 
As a bonus, it would also provide an 
independent confirmation of the emission-line selection.
In Figure~\ref{spec_comp}, we compare the one-dimensional spectra 
formed in the two different ways using two of the emitters. Despite 
the very different procedures used, the two methods produce spectra 
that agree well in shape and normalization.

\begin{inlinefigure}
\includegraphics[width=3.4in]{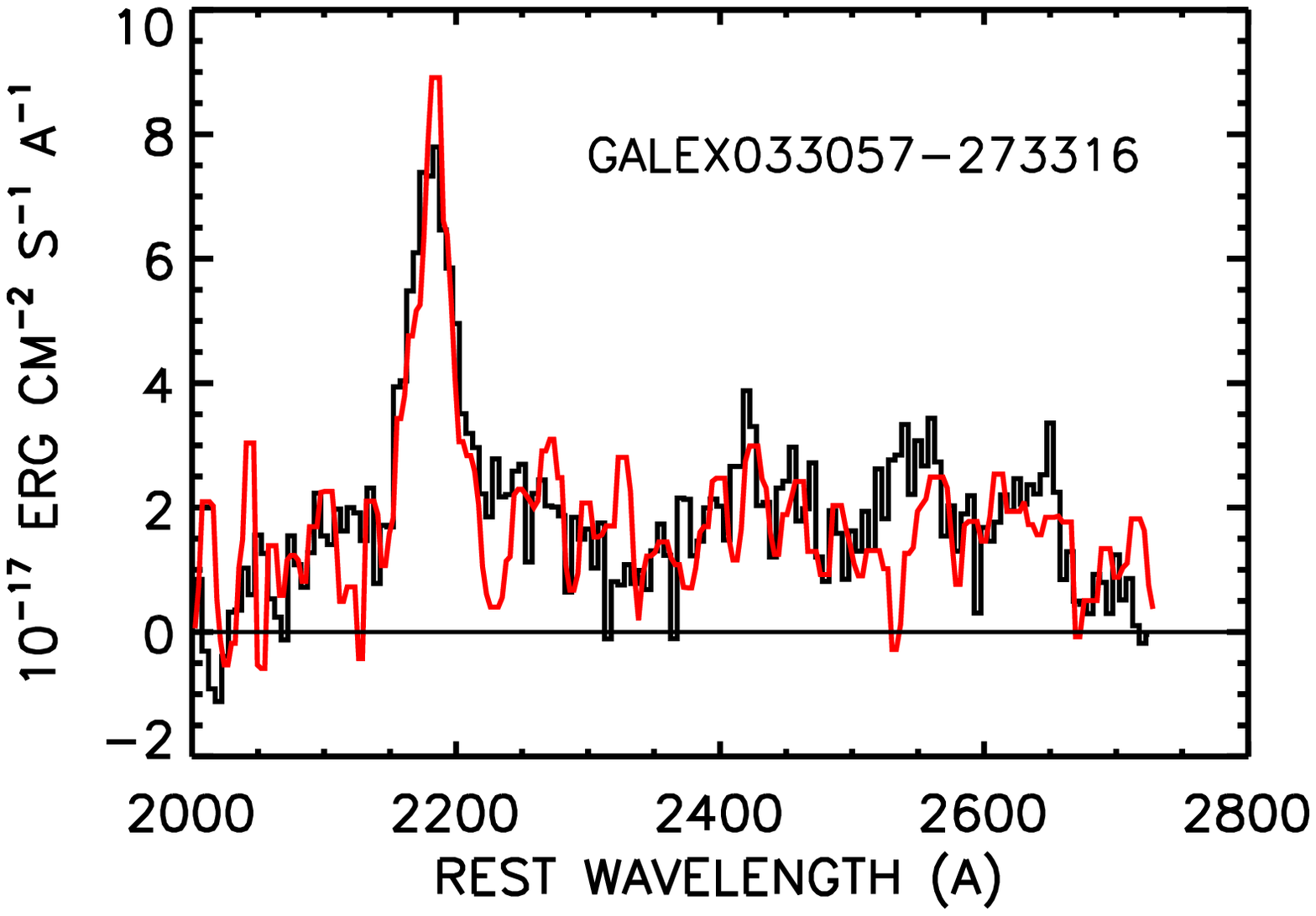}
\includegraphics[width=3.4in]{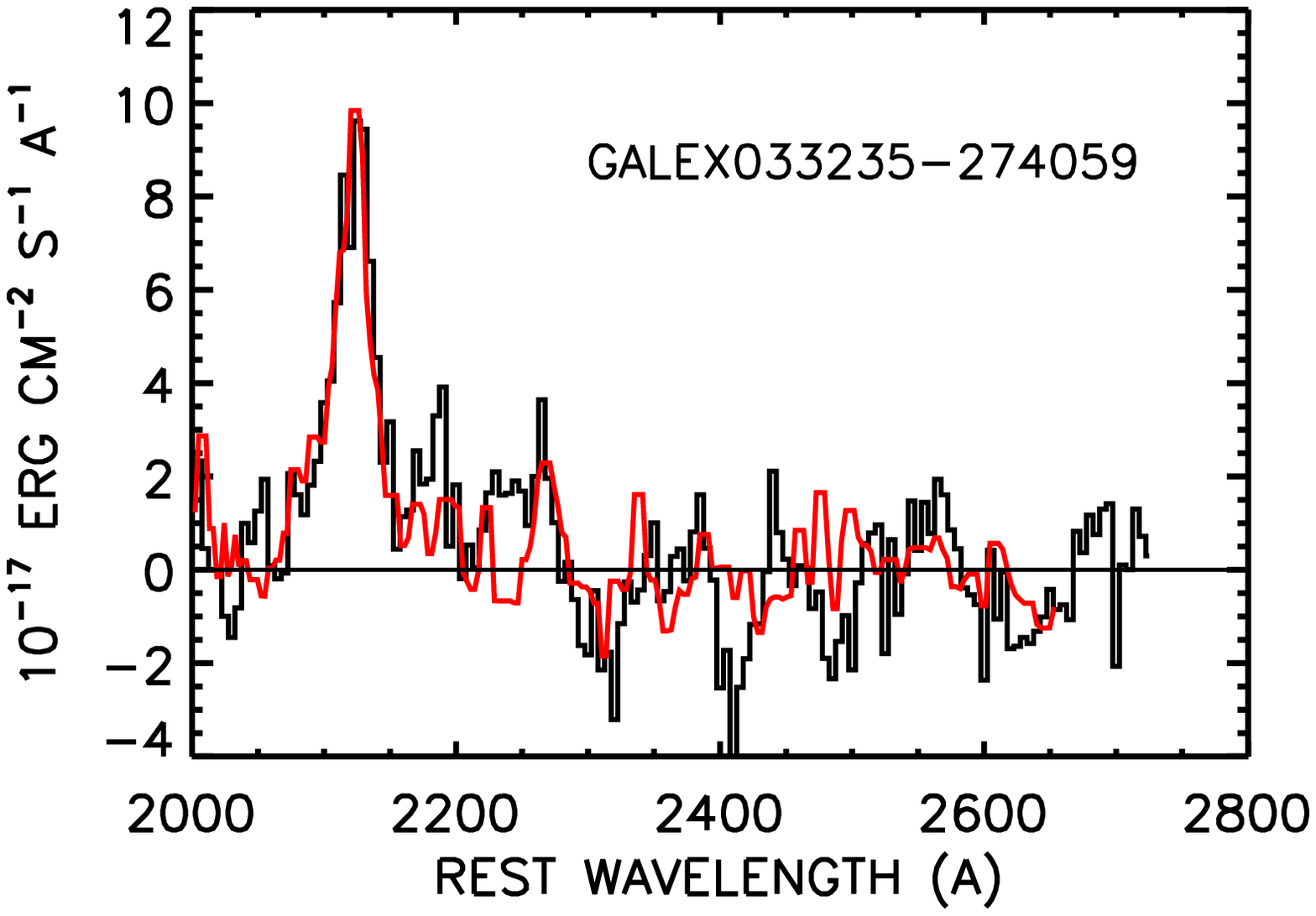}
\caption{
Comparison of the spectra of two emitters in the CDFS-00 obtained 
in two ways:  from our three-dimensional NUV data cube (black) and 
from our version of the {\em GALEX\/} pipeline (red). For each 
emitter, the name is given in the upper right of the panel. As
discussed in the text, the extraction methods and the absolute
flux calibration differ for the data cube spectra and the pipeline 
spectra, so the agreement provides a test of the two methods.
\label{spec_comp}
}
\end{inlinefigure}

For each of the 28 objects, we checked that the discovered line was 
not clearly another line rather than \lya\ by using the combined
FUV and NUV spectra. 
For the spectrum shown in Figure~\ref{galex}, the discovered line 
was not \lya, but as we noted above, this object was from the
pipeline sample rather than from our data cube sample.
All of the lines in our data cube sample appear to be \lya.

We measured the redshifts, the \lya\ fluxes, and the line widths
by fitting a Gaussian to the primary line in each spectrum. 
Since, in general, 
the continua are faint in the spectra, we obtained the EWs of the 
lines by dividing the measured \lya\ flux by the continuum flux 
measured from the broadband NUV continuum image. 
We also made a classification of 
whether the emitter was definitely an AGN based on the
presence of high-excitation emission lines in its UV spectrum, or 
whether it could be a star-forming galaxy (see, e.g., Cowie et al.\ 2010). 
Our candidate star-forming galaxy sample will still contain some remaining 
AGNs, so optical spectra and/or X-ray imaging data are needed to 
make a final determination of whether a galaxy is, in fact, 
star-forming.

\subsection{Completeness of Recovery Versus Flux}
\label{comp}

Because of the low spatial and spectral resolution of the {\em GALEX\/}
grism data, the emitters almost all appear as unresolved 
objects in the narrowband slices or in the two-dimensional
spectral images (Figure~\ref{spectralimages}). The one
exception to this is GALEX033246-274154, which is clearly
extended. In Figure~\ref{slice_images}a, we show a contoured
image (black) of a typical emitter, 
and in Figure~\ref{slice_images}b,
we show a similar image for GALEX033246-274154. 
In both panels, we also show the point spread function contours
formed from the average of the three brightest sources in the 
sample (red). 
There is a clear distinction between the two panels.
In order to be spatially resolved, the object must be very extended. 
Thus, we shall interpret GALEX033246-274154 as a \lya\ blob in the 
discussion section.

\begin{inlinefigure}
\includegraphics[width=3.4in,angle=0]{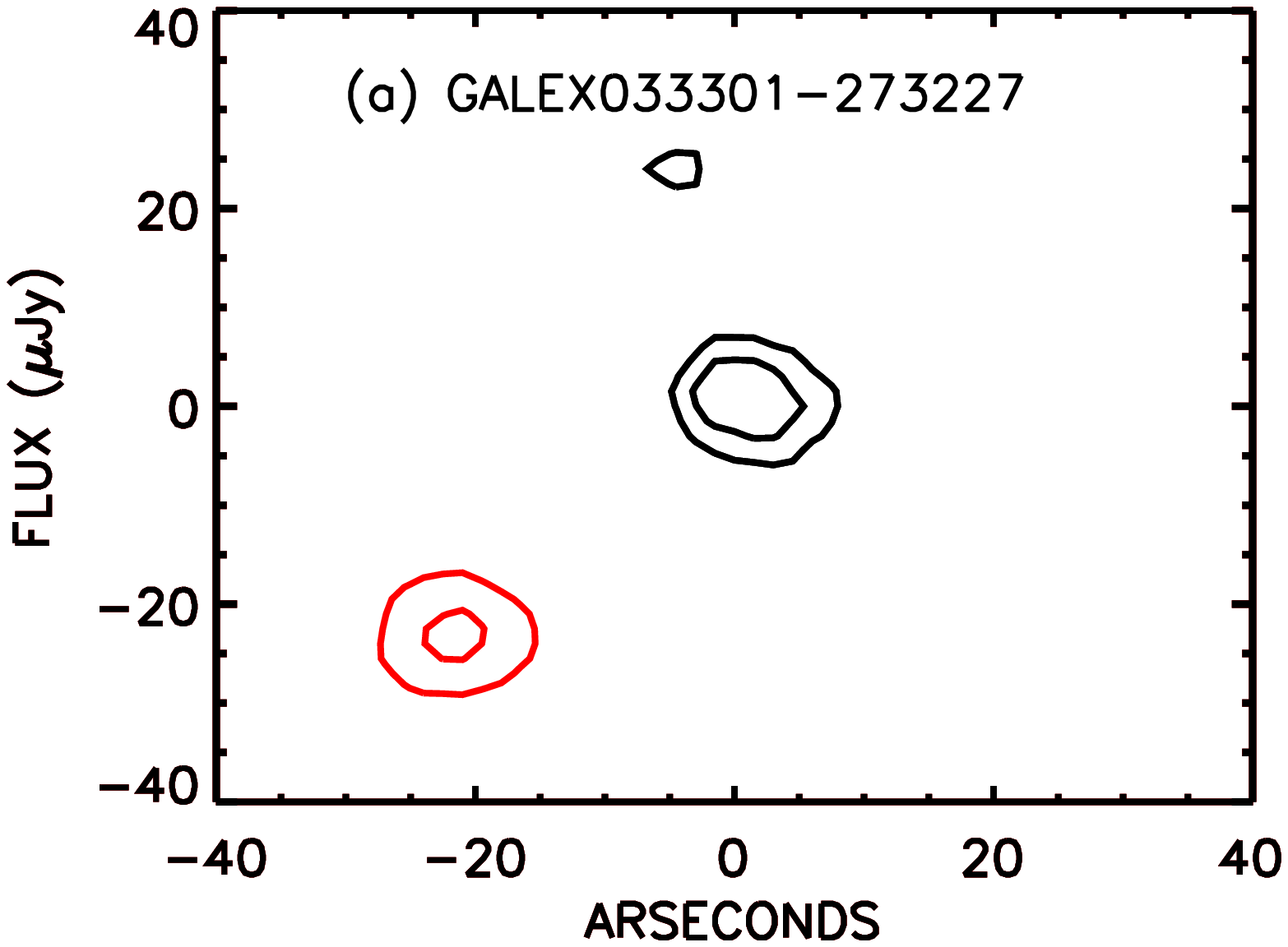}
\includegraphics[width=3.4in,angle=0]{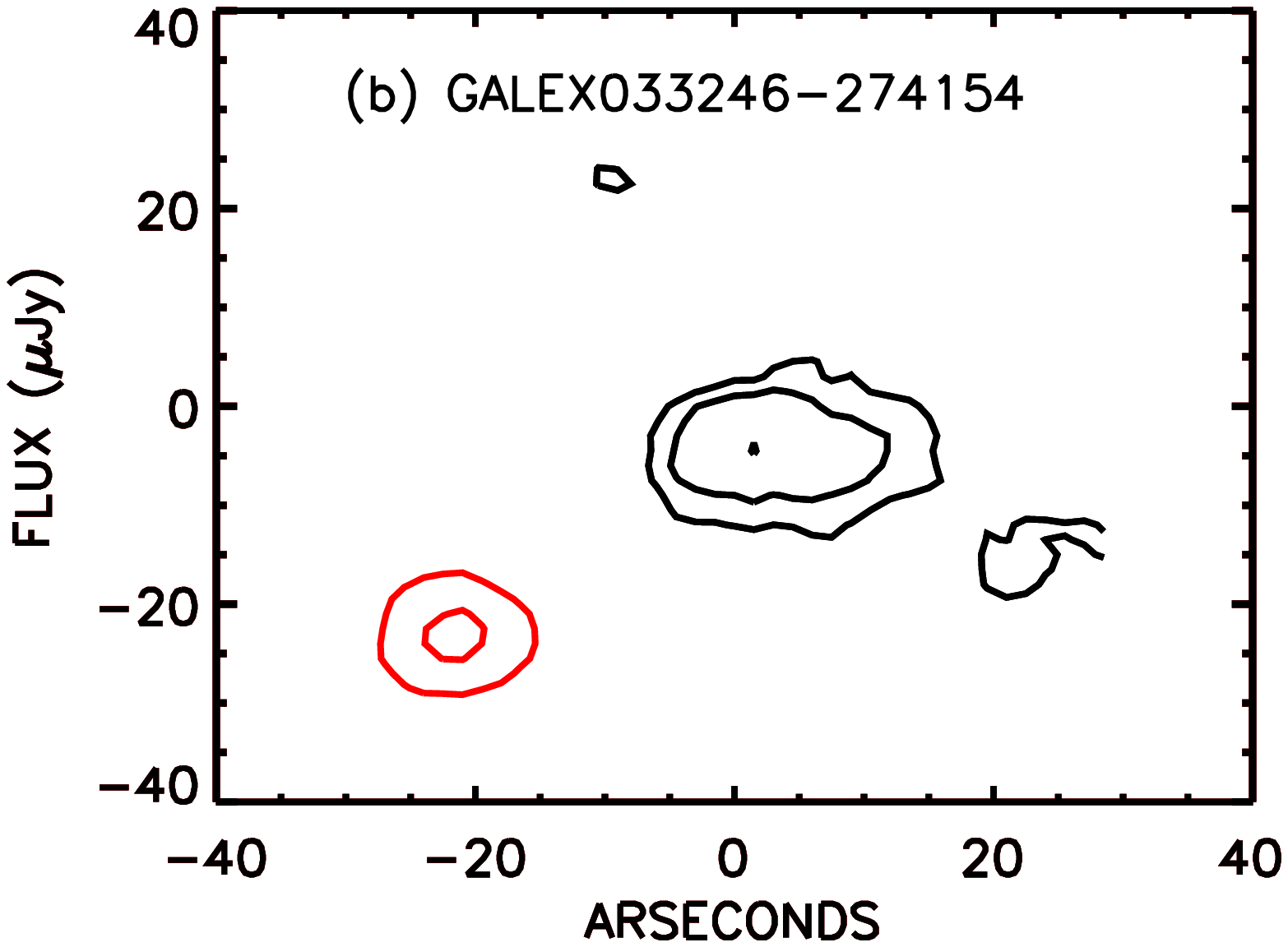}
\caption{
Narrowband contoured images (black) of (a) a typical emitter,
which is spatially unresolved, and (b) the \lya\ blob.
The red contours show the point spread function formed from the 
average of the three brightest sources in the sample. 
\label{slice_images}
}
\end{inlinefigure}

Since nearly all the emitters in our sample are unresolved, it is 
relatively straightforward to estimate the completeness with which we
can detect emitters of a given flux, since we do not 
need to consider morphological or size differences.
In addition, since the observed-frame equivalent widths of the
objects are large compared to the spectral resolution, 
the emission-line peaks have a high contrast with the continuum level 
(Figure~\ref{spect}).
Thus, the selection depends only on the emission-line flux and not on the
continuum level, or, as a corollary, on the equivalent width.
  
In order to estimate the completeness, we added in 3000 objects 
(one hundred emitters in each of the thirty narrowband slices)
of a specified flux with shapes corresponding to the spatial point 
spread function measured using bright objects in the data cube.
The emitters were 
added into the data cube at random spatial positions and uniformly 
distributed through the narrowband slices. We then ran our standard 
selection procedure and found the number of recovered objects. 
At fluxes above $10^{-15}$~erg~cm$^{-2}$~s$^{-1}$, our selection 
is more than $80\%$ complete, but it drops extremely rapidly below this 
value (see Figure~\ref{complete}). We shall use this in
calculating the \lya\ luminosity function.

\begin{inlinefigure}
\includegraphics[width=2.8in,angle=90]{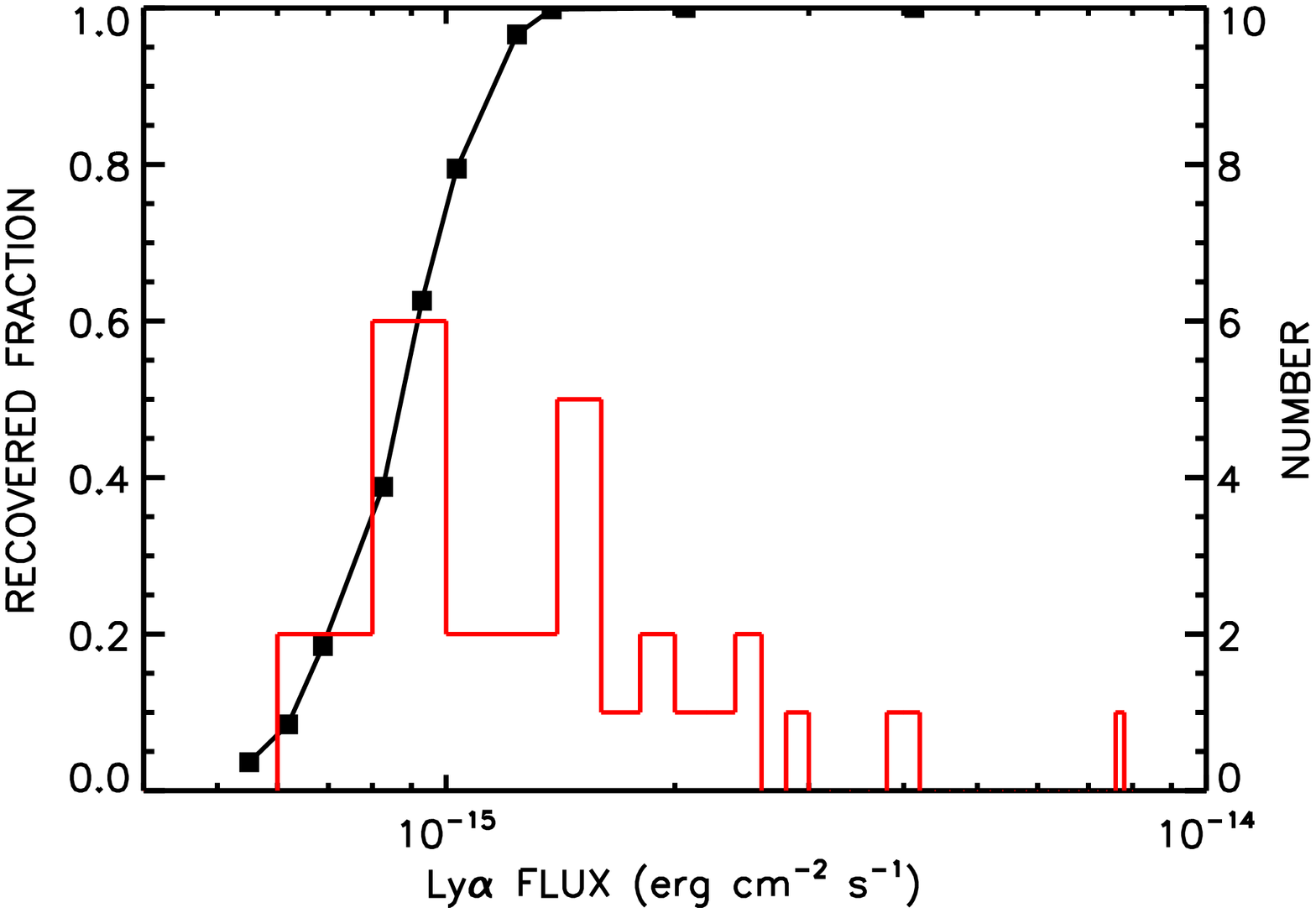}
\vskip -0.4cm
\caption{
Fraction of simulated sources recovered as a function of 
the emission-line flux (black curve with squares at
the measured fluxes). The red histogram shows the number of
detected sources as a function of flux in 
$2\times10^{-16}$~erg~cm$^{-2}$~s$^{-1}$ bins. The counts 
flatten out and drop at low fluxes where 
the incompleteness becomes important.
\label{complete}
}
\end{inlinefigure}

\section{New \lya\ Emitters in the CDFS}
\label{newLAE}

In Table~\ref{faint_z1_sample_table}, we summarize the properties
of the 28 new $z\sim1$ LAEs in the CDFS-00 found in our
data cube search. These are all objects that were not
present in the pipeline sample.
In Column~1, we give the {\em GALEX\/} name; in Columns~2 and 3,
the J2000 right ascension and declination; in Columns~4 and 5, 
the NUV and FUV AB magnitudes; in Column~6, the redshift from the 
{\em GALEX\/} UV spectrum; in Column~7, the \lya\ fluxes
with $1\sigma$ errors; in Column~8, the logarithm of the \lya\ 
luminosity; in Column~9, the rest-frame EW(\lya) with
$1\sigma$ errors; in Column~10, an entry of ``AGN'', if the source 
is classified as an AGN based on its UV spectrum; in Column~11,
the logarithm of the $2-8$~keV flux, if the source is detected
in the Extended CDFS, or an entry of ``ECDFS'', if the source is 
in the ECDFS region but is not detected; in Column~12, the 
optical ground-based redshift; and in Column~13, the optical 
spectral classification. 

We measured the NUV and FUV AB magnitudes in $8''$ diameter apertures 
centered on the emitter positions. We corrected these to total 
magnitudes by measuring for the brighter objects ($20-23$~mag range) 
the offset between the aperture magnitudes and the magnitudes of
the pipeline sample.
The $1\sigma$ errors obtained by measuring magnitudes at
random positions in the images are 26.0 for the NUV and 26.7
for the FUV. All of the objects are bright in the NUV, but some
are not detected in the FUV. For objects where we measure
a negative flux in the FUV, the quantity in Column~5 is the magnitude
corresponding to the absolute value of the flux with a minus
sign in front to indicate that the flux was negative.

When allowance is made for the masking done on 
bright sources (see Sections~\ref{constructdatacubes} 
and \ref{searchdatacubes}),
the area covered is 2286~arcmin$^2$, and the 
data cube emission-line search procedure covers a redshift
range for \lya\ detections of $z=0.67-1.16$. The observed
comoving volume is $2.3\times10^{6}$ Mpc$^{3}$.

Thirteen of the LAEs lie in the area covered by the deep
X-ray exposures of the ECDFS (Lehmer et al.\ 2005; Virani et al.\ 2006);
six of these thirteen LAEs are detected in X-rays (see Column~11).
At $z\sim1$, the ECDFS sensitivity limit 
($f_{2-8~{\rm keV}}\sim 6.7\times 10^{-16}$~erg~cm$^{-2}$~s$^{-1}$) 
is close to the X-ray luminosity
threshold of $10^{42}$~erg~s$^{-1}$ that is usually used to
define AGN activity (e.g., Hornschemeier et al.\ 2001; Barger et al.\ 2002;
Szokoly et al.\ 2004; Silverman et al.\ 2005),
so the sources that are not detected have X-ray luminosities 
consistent with being star-forming galaxies. 
We hereafter refer to the sources that are X-ray detected (and
hence are clearly AGNs based on their X-ray luminosities) as X-ray AGNs.

Nine of the LAEs have optical spectroscopic redshifts 
(see Column~12). For all of these sources, the optical
redshift confirms the UV redshift derived from the data cube search.
Four of these nine are X-ray AGNs, which is not too surprising,
given that the spectroscopy in the ECDFS region was concentrated on 
X-ray selected objects. The mean offset and dispersion between the
ground-based optical and the {\em GALEX\/}
UV redshifts is $z_{\rm ground} -  z_{\rm galex} =-0.004 \pm 0.003$. 

All three of the sources in the ECDFS region that were classified as 
AGNs based on the presence of high-excitation lines in their UV spectra 
are also X-ray AGNs. In the other direction, three of the 6 sources
classified as X-ray AGNs 
were already classified as AGNs based on their UV spectra.
Of the remaining three sources, GALEX033131-273429 is classified as
a Seyfert~2 (Sy2) based on the presence of [NeV]$\lambda3426$ in 
its optical spectrum, and, as can be seen in 
Figure~\ref{spect}a, has a single narrow but very strong \lya\ 
emission line in the UV but no obvious CIV$\lambda1549$ emission;
GALEX033202-274319 is 
classified as a broad-line AGN from its optical spectrum; and
GALEX033335-273934 has not been optically observed. In the case of the 
latter two objects, their redshifts are poorly positioned to detect 
either OVI$\lambda 1036$ or CIV$\lambda1549$ in their UV spectra, 
and only their \lya\ lines are seen.  

Classifying sources as AGNs based on their UV spectra can be challenging,
since quite often CIV is not accessible, and, as can be seen from 
Figure~\ref{spect}, OVI is typically quite weak.
The fraction of LAEs in the ECDFS region that were not classified as AGNs 
based on their UV spectra (10 sources) but are classified as AGNs based 
on their optical spectra (2 sources) is 20\%.  This is very similar to 
the $\sim 20$\% found for the LAEs in the {\em GALEX\/} pipeline 
samples (Finkelstein et al.\ 2009b; Scarlata et al.\ 2009; 
Cowie et al.\ 2010; see Cowie et al.\ 2011 for a discussion of the 
variations in the measurements of this quantity). 

We conclude that the bulk of the LAEs found in our data cube search
are real and that the UV spectroscopic redshifts based on the
\lya\ identifications are reliable. 

\begin{inlinefigure}
\includegraphics[width=2.8in,angle=90]{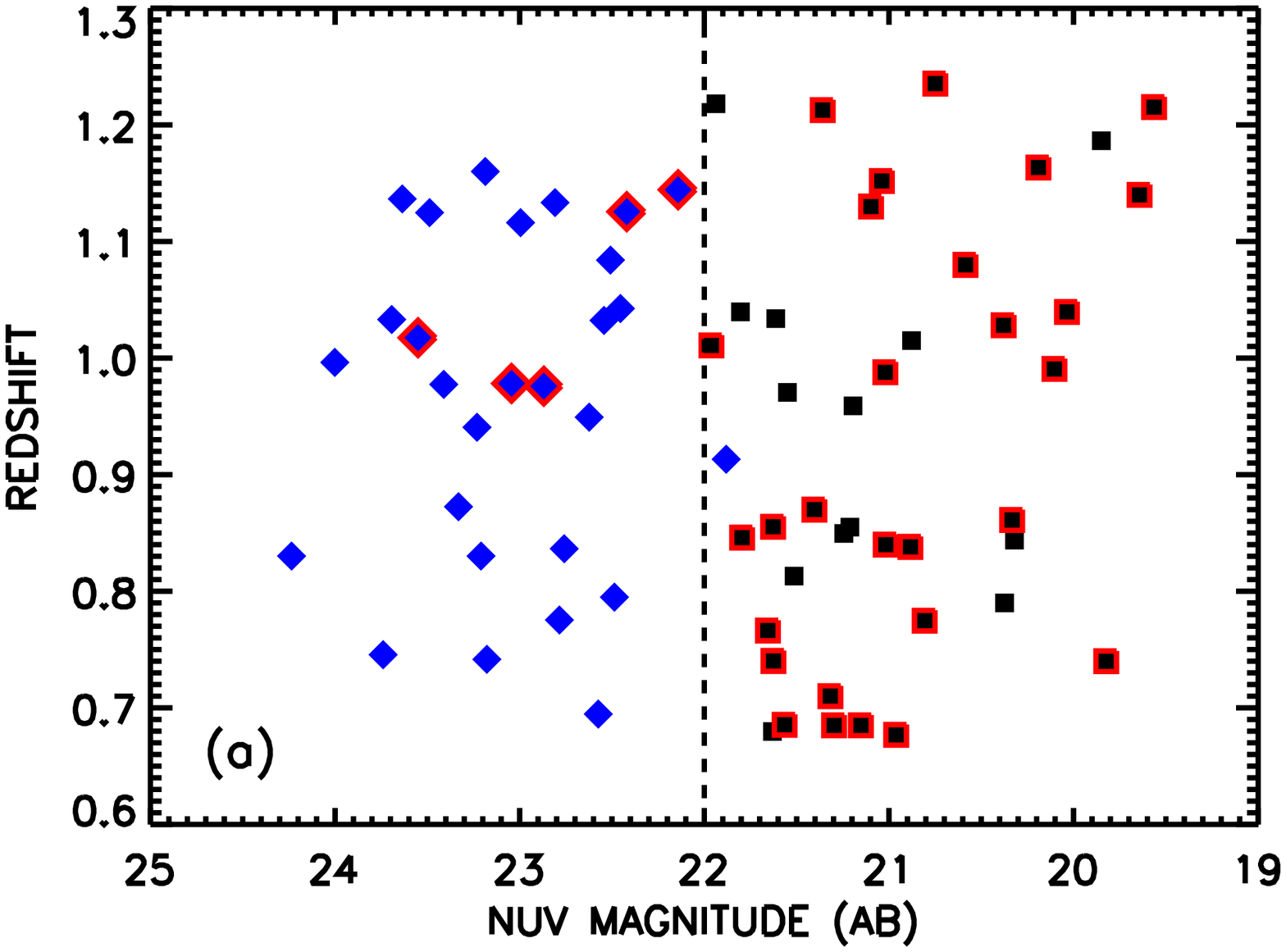}
\vskip -0.5cm
\includegraphics[width=2.8in,angle=90]{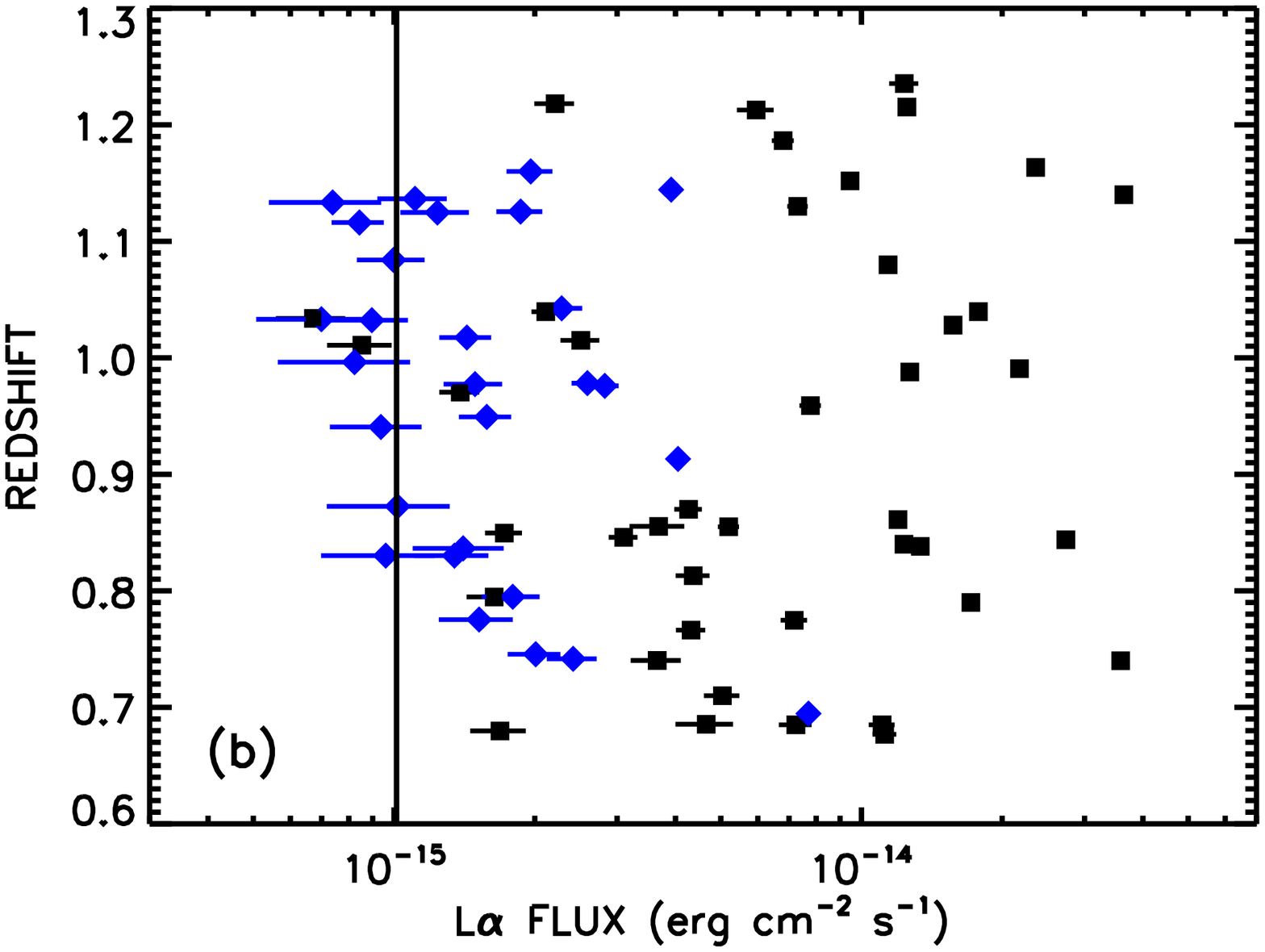}
\vskip -0.5cm
\caption{
(a) {\em GALEX\/} redshift vs. NUV magnitude.
Black squares show the CDFS-00 LAEs found by Cowie et al.\ (2010, 2011)
in the {\em GALEX\/} pipeline sample of NUV$<22$ 
(dashed line) continuum-selected objects in the CDFS-00 field.
Blue diamonds show the new 
CDFS-00 LAEs found in our data cube search; none of these 
were in Cowie et al.
Note that the redshift range searched in the data cube is slightly 
smaller. Objects with high-excitation UV lines (we classified these
sources as AGNs based on their UV spectra) are surrounded by larger 
red symbols.
(b) {\em GALEX\/} redshift vs. \lya\ flux. The symbols are as in (a).
We show $1\sigma$ error bars on the \lya\ fluxes.  
Solid line shows the \lya\ flux above which more than
$80\%$ of the emitters are detected (see Fig.~\ref{complete}).
\label{z_mag}
}
\end{inlinefigure}

\section{Discussion}
\label{disc}

We now consider how the properties of the CDFS-00 LAEs in our data 
cube sample compare with those of the CDFS-00 LAEs in the pipeline 
sample. In Figure~\ref{z_mag}a, we illustrate how, by construction, 
the pipeline misses sources fainter than the pipeline magnitude 
limit of NUV$\sim22$ (dashed line). 
We show the {\em GALEX\/} redshifts versus the NUV magnitudes 
for the CDFS-00 LAEs found by Cowie et al.\ (2010, 2011)
(black squares) and the new CDFS-00 LAEs found by our NUV 
data cube search (blue diamonds).
In contrast, in Figure~\ref{z_mag}b (symbols and colors as in
Figure~\ref{z_mag}a), we show how there is substantial 
overlap in the \lya\ emission-line fluxes of the two samples,
since this quantity depends on both the NUV magnitude and the 
EW(\lya).  Here we can
see that the pipeline extraction begins to miss sources below
a \lya\ flux of about $8\times10^{-15}$~erg~cm$^{-2}$~s$^{-1}$
and becomes progressively more incomplete at fainter fluxes,
until below $2\times10^{-15}$~erg~cm$^{-2}$~s$^{-1}$, very few
sources are detected in the pipeline extraction. 

In Figure~\ref{z_mag}a, we use larger red symbols to denote 
sources with high-excitation UV lines (in addition to \lya) in 
their {\em GALEX\/} spectra. These sources are classified as AGNs 
based on their UV spectra.  We can see that at NUV$<22$, most of
the sources are AGNs.  Thus, the only way to probe the bulk of
the LAE galaxy population at these redshifts is to do a data
cube search of the present type.

The sources found in the data cube search and missed
in the pipeline are those with high EWs. The
progressive increase in incompleteness with decreasing 
\lya\ flux can be illustrated by the number of sources lying 
above the EW threshold at a given \lya\ flux. We show
this in Figure~\ref{obsEW} (symbols and colors as in
Figure~\ref{z_mag}), where we plot rest-frame
EW versus \lya\ flux. For a given
\lya\ flux and redshift, the NUV$=22$ continuum cut corresponds
to a given EW. We show these cuts in the figure
at the upper (dashed diagonal) and lower (solid diagonal)
end of the $z=0.67-1.16$ redshift range.
They become progressively lower as we move
to fainter \lya\ fluxes, such that fewer of the emitters are 
included in the pipeline sample.

\begin{inlinefigure}
\includegraphics[width=2.8in,angle=90]{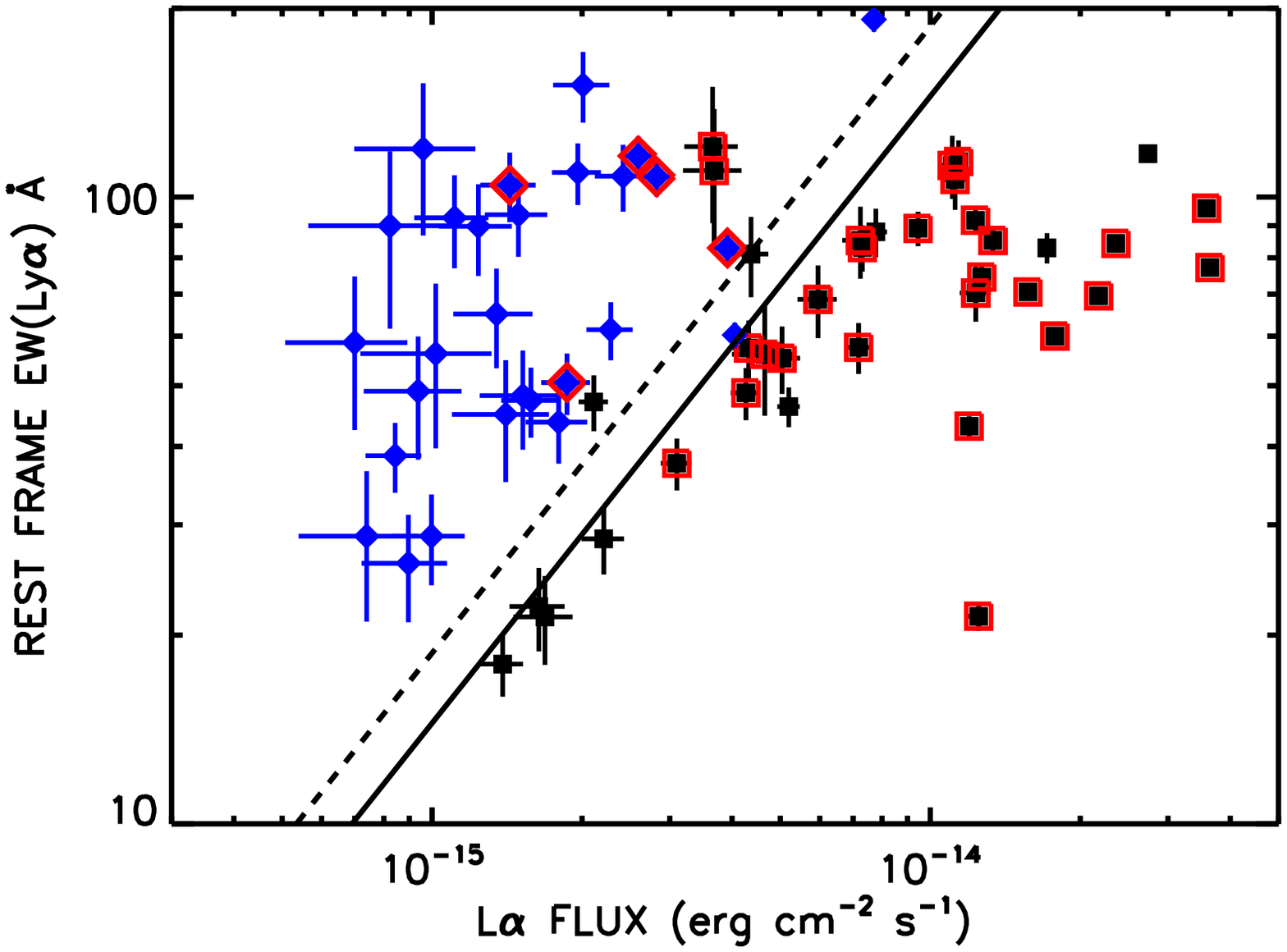}
\vskip -0.4cm
\caption{
Rest-frame EW(\lya) vs. \lya\ flux. Black squares show the CDFS-00 
LAEs found by Cowie et al.\ (2010, 2011) in the
{\em GALEX\/} pipeline sample of NUV$<22$ continuum-selected objects
in the CDFS-00 field.
Blue diamonds show the new CDFS-00 LAEs found in our
data cube search; none of these were in Cowie et al.
Objects with high-excitation UV lines (we classified these
sources as AGNs based on their UV spectra) are surrounded by 
larger red symbols. We show $1\sigma$ error bars on the \lya\ 
fluxes and equivalent widths. The {\em GALEX\/} pipeline sample
only has sources with EWs below the value set by the
NUV magnitude limit (diagonal lines illustrate the effects of
the redshift range: $z=0.67$---solid; $z=1.16$---dashed).
The black squares should lie below these lines,
and the blue diamonds should lie above these lines.
Clearly, as one moves to lower \lya\ fluxes, the continuum-selected 
pipeline extraction starts to miss most of the LAEs.
\label{obsEW}
}
\end{inlinefigure}

We next inspected the UV/optical images of the full CDFS-00 LAE 
sample (i.e., this includes both the pipeline and the data cube 
LAE samples) in the redshift range $z=0.67-1.12$. All but one
of these LAEs are covered by deep $U$-band imaging data
obtained with the wide-field (1~deg$^2$ field-of-view) imaging
camera MegaPrime/MegaCam on the 3.6~m Canada-France-Hawaii Telescope (CFHT). 
The Canadian Astronomy Data Centre (CADC) reduced the data 
using the MegaCam Image Stacking Pipeline (MegaPipe; Gwyn 2008)
and give the $5\sigma$ AB magnitude limits as ranging from 
26.5 to 27.0.  Only one of the observed objects (GALEX033246-274154) 
does not have a clear counterpart in the CFHT $U$-band.

\begin{inlinefigure}
\vskip 0.4cm
\includegraphics[width=3.0in]{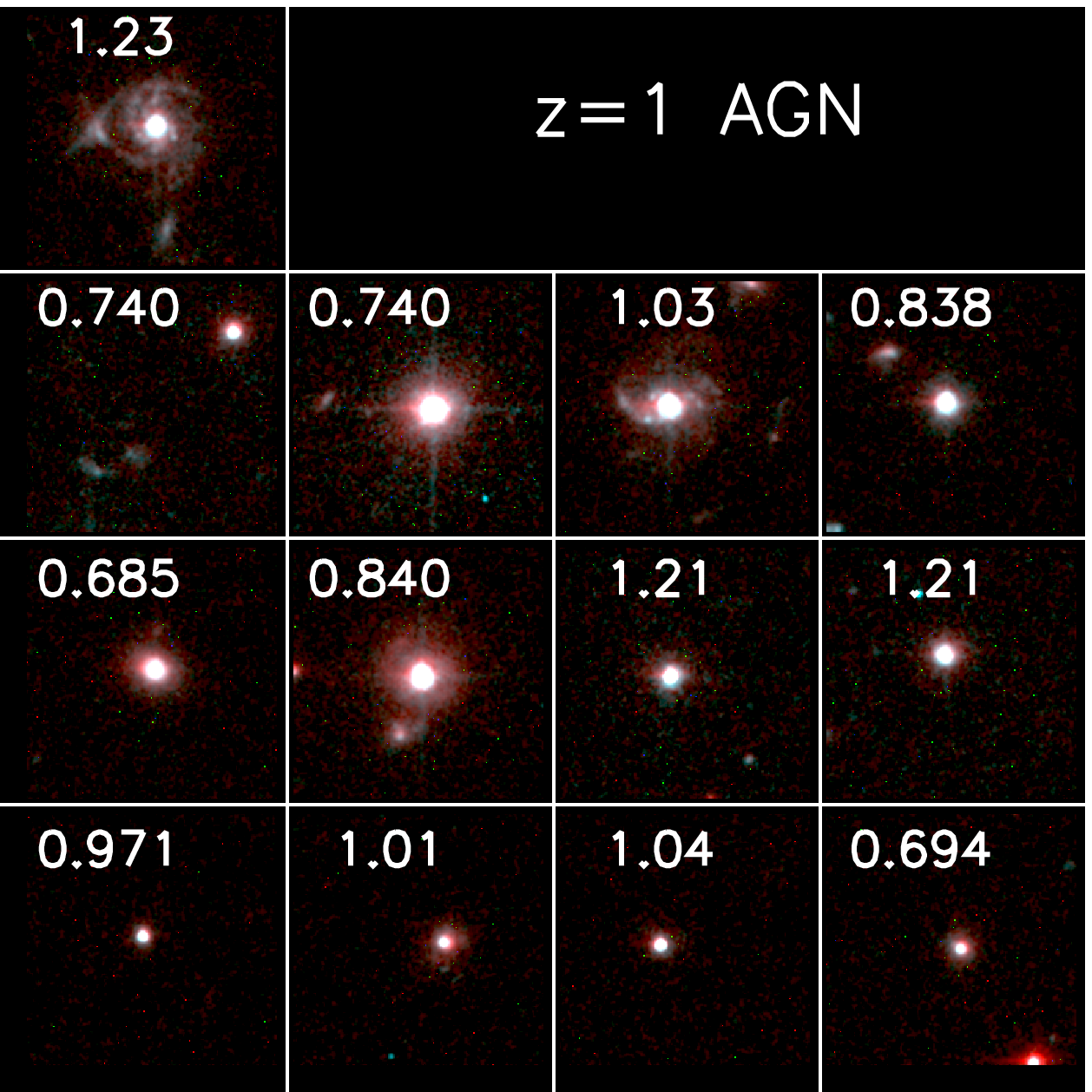}
\includegraphics[width=3.0in]{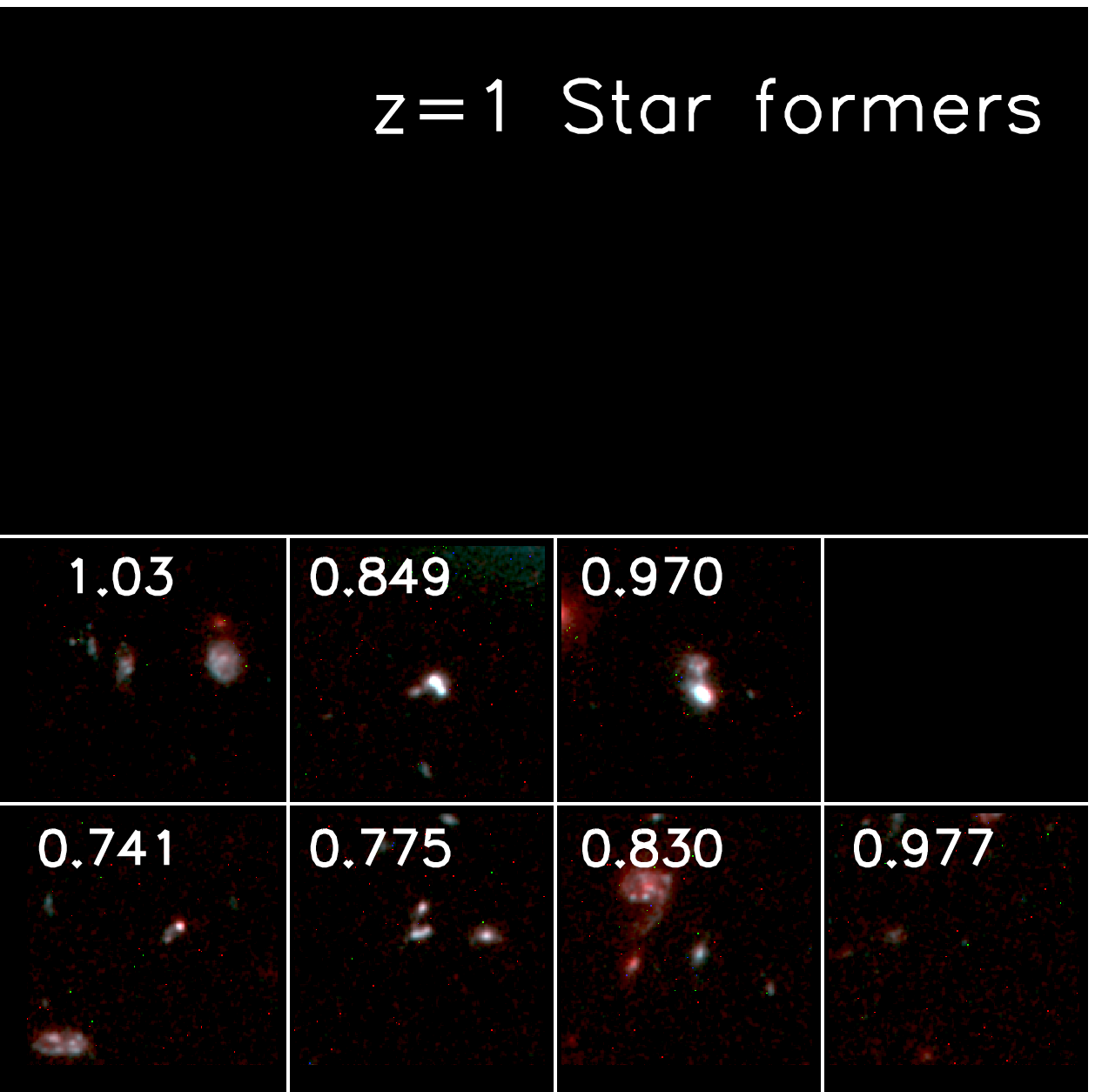}
\caption{Images from the GEMS survey (Rix et al.\ 2004)
of LAEs found either in the CDFS-00 {\em GALEX\/} pipeline 
sample or in our data cube search. 
The lower 7 sources are candidate star-forming galaxies in the region, 
and the top 13 sources are AGNs based on either their UV continuum 
lines or their X-ray luminosities.  Each thumbnail is
$6''$ on a side. The blue and green colors correspond to the
F606W GEMS image, and the red color corresponds to the F850W
GEMS image.
\label{gems_thumbs}
}
\end{inlinefigure}

In addition, 20 of the 70 LAEs in the full CDFS-00 sample
are in the GEMS region (Rix et al.\ 2004), 
where high spatial resolution {\em HST\/} images are available. 
We show the GEMS images in 
Figure~\ref{gems_thumbs}. We have divided these into two groups: 
AGNs (based on either their UV spectral signatures or their X-ray 
luminosities, which turns out to include all of the objects that were 
classified as AGNs based on their optical spectra) 
and star-forming galaxies (no high-excitation UV lines and not 
detected in X-rays). 
It appears that the morphologies
of the objects follow their classifications: the AGNs
are dominated by compact nuclei, while the star-forming galaxies 
are primarily resolved small blue galaxies, often with multiple 
components.

In addition to not being detected in the CFHT $U$-band,
GALEX033246-274154 (last thumbnail of the bottom row in 
Figure~\ref{gems_thumbs}) is not 
clearly detected in the {\em HST\/} images.
GALEX033246-274154 appears to be a rather unusual object.
It is clearly visible both in the NUV continuum image
and in the UV spectral image (Figure~\ref{spectralimages}) and hence 
is real.  Furthermore, it is spatially resolved 
(Figure~\ref{slice_images}b).
Its major axis is $18''$ and its minor axis is about
$9''$, giving a linear major-axis diameter of 120~kpc and placing
it in the category of the giant \lya\ blobs, though its
\lya\ luminosity is at the low end of these objects
(Matsuda et al.\ 2011). 

\begin{figure*}
\centerline{\includegraphics[width=1.7in,angle=90]{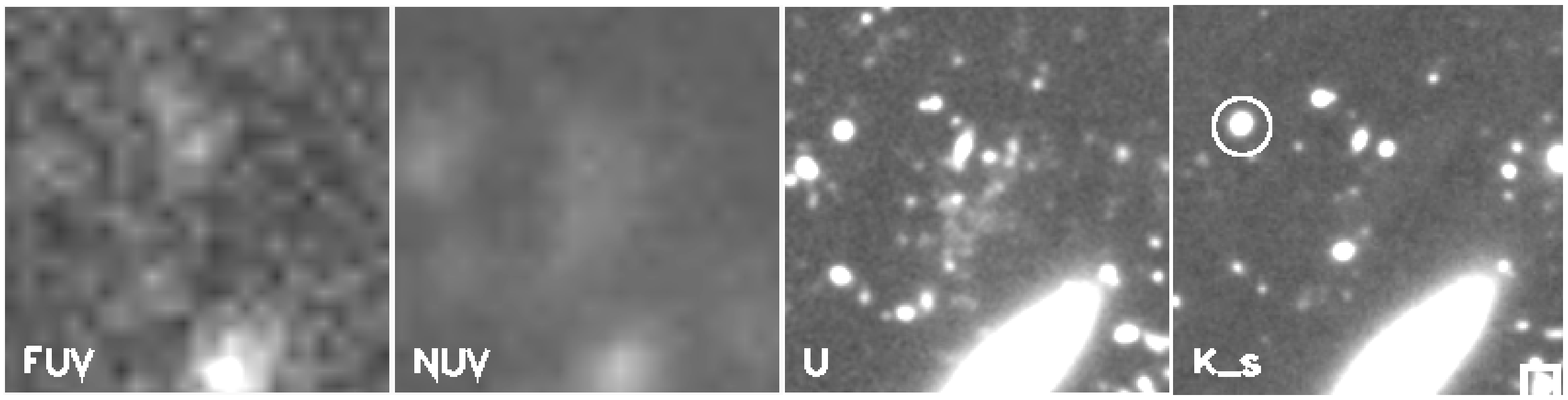}}
\caption{Images of the giant \lya\ blob GALEX033246-274154. From left 
to right the panels show the FUV, NUV, VIMOS $U$ (Nonino et al.\ 2009), 
and $K_s$ images.
Each thumbnail is $37\farcs5$ on a side. 
In the final panel, we circle the nearest ($14''$)
X-ray source in the ECDFS sample to the blob. 
This AGN is at $z=0.740$ 
and hence cannot be the excitation source for the blob.  We also mark 
with a square an AGN located $25''$ from the blob, whose redshift
of $z=0.982$ roughly matches that of the blob ($z=0.977$).
\label{blob}
}
\end{figure*}

\begin{figure*}
\centerline{\includegraphics[width=5in,angle=90]{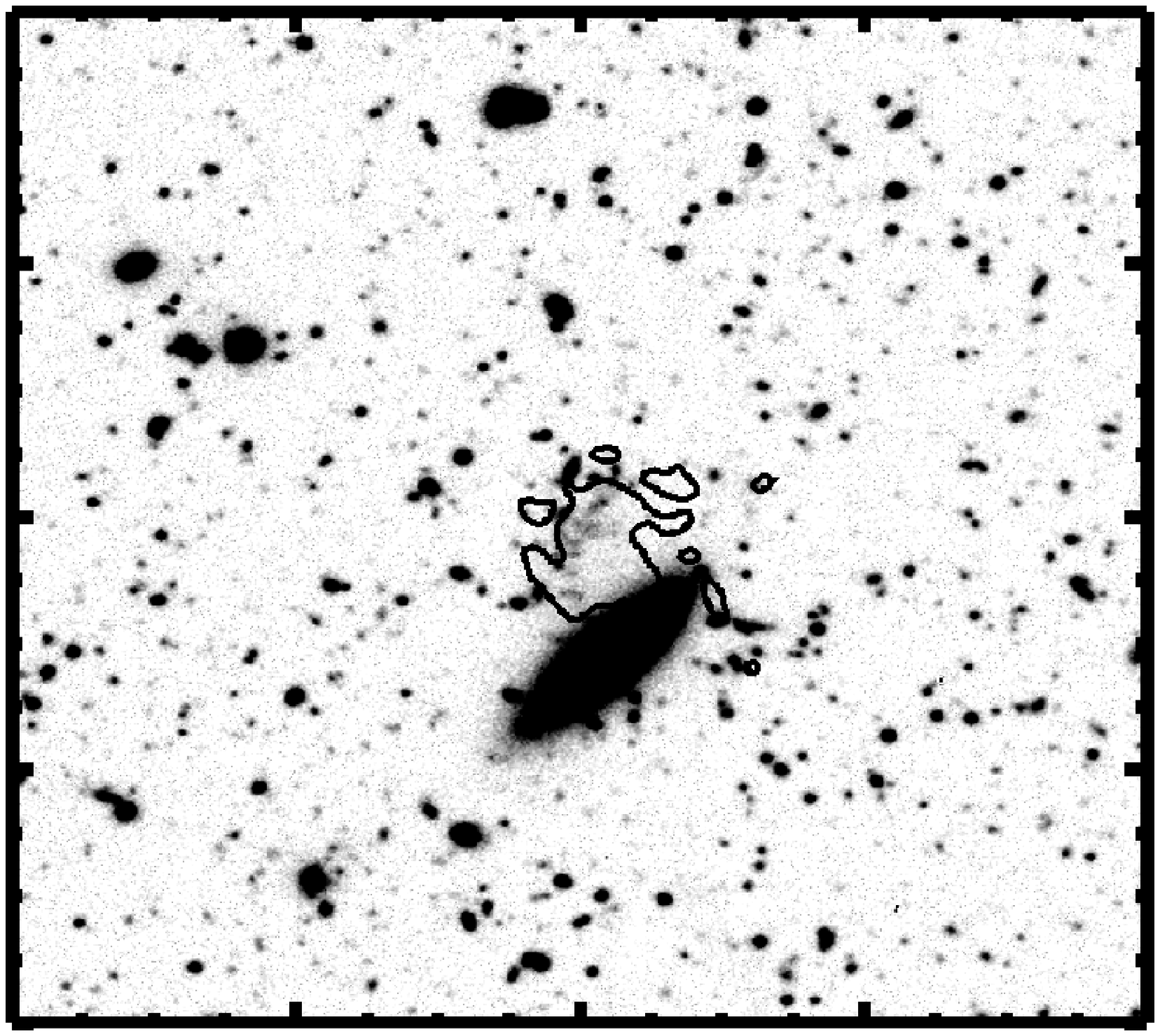}}
\vskip -0.2cm
\caption{Contour of the \lya\ emission-line region of the
giant \lya\ blob
GALEX033246-274154 superimposed on the ultradeep VIMOS $U$-band
exposure of the area (Nonino et al.\ 2009). The
emitter corresponds to an area of diffuse though structured
continuum emission. Each large tick on the axes corresponds
to an angular size of $30''$ or a linear size of 240~kpc
at the $z=0.977$ redshift of the object.
\label{new_blob}
}
\end{figure*}

Fortunately, a much deeper $U$-band image than the CFHT MegaPrime
image was taken (Nonino et al.\ 2009) with the VIMOS instrument on the 
ESO VLT and covers the central region of the field, 
including this object.
In Figure~\ref{blob}, we compare the FUV image (1st thumbnail) of
the source (located at the center of each image) and the NUV image
(2nd thumbnail) with the VIMOS $U$-band image (3rd thumbnail) and 
a $K_s$-band image (final thumbnail). In Figure~\ref{new_blob},
we overlay the contours of the emission-line region on the VIMOS
$U$-band image.
From these images, we see that the emission-line region corresponds
to a faint diffuse $U$-band continuum structure. The object appears
very similar to higher redshift \lya\ blobs 
(e.g., Prescott et al.\ 2012).

We have marked the nearest X-ray selected AGN (circle) on the final 
thumbnail in Figure~\ref{blob}. It lies at about $14''$ from the blob. 
However, this AGN has a redshift of $z=0.740$ and hence cannot be the 
excitation source for the blob. A more probable candidate (square) is 
an AGN that lies $25''$ (170~kpc) from the blob. Its redshift of 
$z=0.982$ roughly matches that of the blob ($z=0.977$).

Matsuda et al.\ (2011) found 10 giant \lya\ blobs at $z=3.1$
in a comoving volume that is almost identical to the present one. 
This suggests that the comoving volume density of giant blobs has 
dropped by almost an order of magnitude between $z=3$ and $z=1$ 
(see also Keel et al.\ 2009). However, a larger sample and a fuller 
understanding of the selection effects in the two samples would be 
needed to refine this result. We postpone a detailed analysis to a
future paper (I.~Wold et al.\ 2012, in preparation), 
where we plan to conduct data cube searches of all 
of the deep {\em GALEX\/} grism fields.

Finally, we computed the \lya\ luminosity function of the CDFS-00
LAE galaxy sample in the redshift range $z=0.67-1.16$ using the 
$1/V$ technique (Felten 1976), calculating the accessible volumes 
from the area of the sample.  We used a flux limit
of $7\times 10^{-16}$~erg~cm$^{-2}$~s$^{-1}$ and only included
sources that were not classified as AGNs in any way
and that have rest-frame EW(\lya)$\ge20$~\AA.
(Note that the EW(\lya)$\ge 20$~\AA\ criterion is what is
normally used to define the high-redshift LAE population; 
e.g., Hu et al.\ 1998.) With these criteria, 
the sample includes two CDFS-00 LAE galaxies from the pipeline
sample of Cowie et al.\ (2011; their Table~2), and the rest come 
from our data cube sample. 
We show the \lya\ luminosity function (black squares) 
in Figure~\ref{high_la_lumfun}, along with
the \lya\ luminosity functions at $z=0.3$ 
(blue dotted curve) and $z=3.1$ (red solid curve) using the 
Schechter (1976) function fits given in Cowie et al.\ (2010)
and Gronwall et al.\ (2007), respectively. 
The open squares
show the computation when we do not correct for the effects 
of incompleteness, and the solid squares when we do
using the results from the simulations (Figure~\ref{complete}). 
The dynamic range in luminosity
is not large enough to justify fitting a Schechter function,
but as we illustrate with the red dashed curve, the data can
be well described by the Gronwall et al.\ (2007) luminosity 
function with the normalization ($\phi_\star$) reduced by a factor
of 30.

\begin{inlinefigure}
\plotone{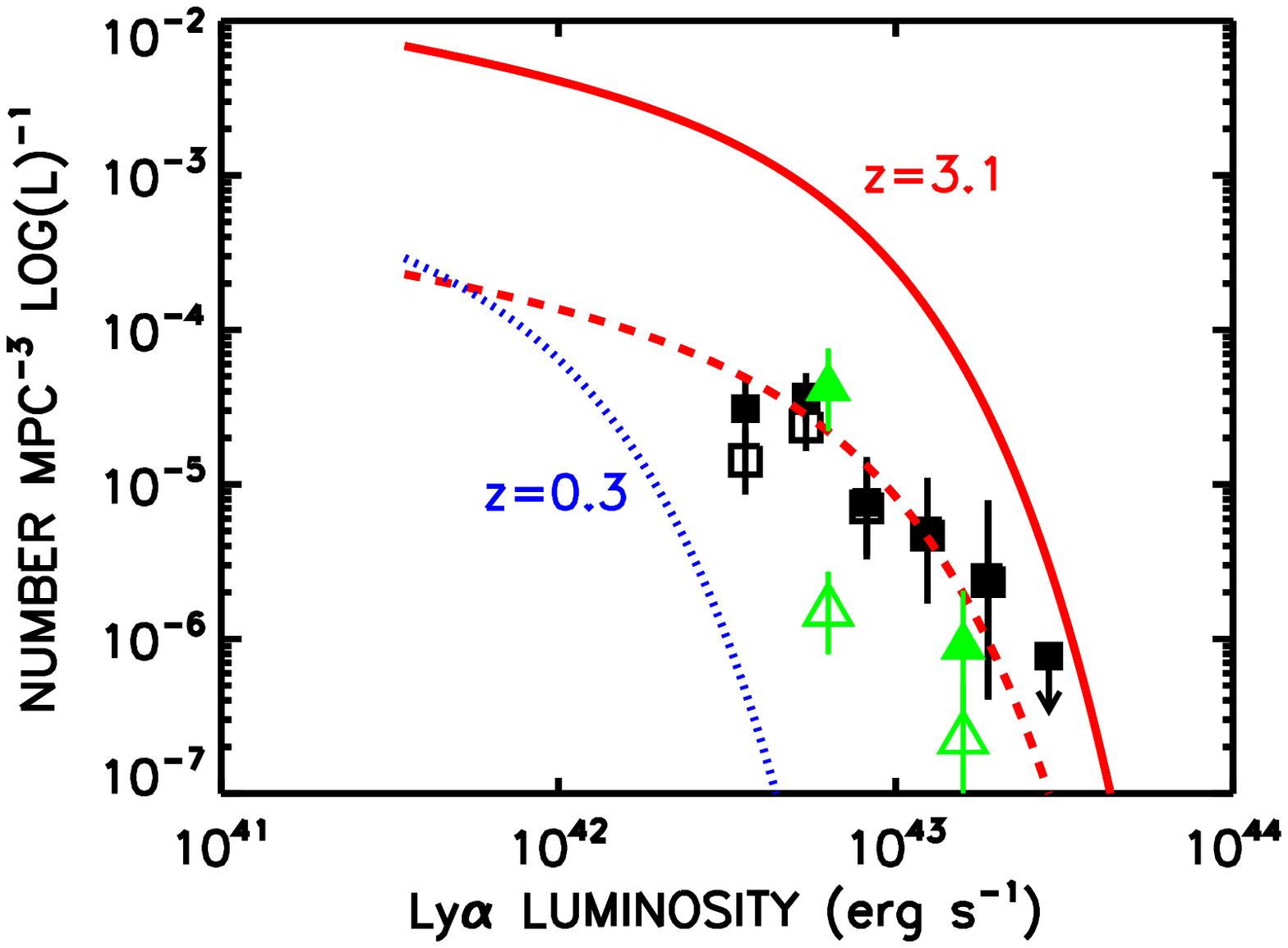}
\caption{Derived \lya\ luminosity function at
$z=0.67-1.16$ for the LAE galaxies in the CDFS-00 with
EW(\lya)$\ge20$~\AA\ from both the 
{\em GALEX\/} pipeline and data cube samples (black squares:
open---raw data; solid---corrected for the effects of incompleteness 
using the results from the simulations; see Figure~\ref{complete}).
Error bars are $\pm1\sigma$ from the Poisson
errors corresponding to the number of sources in the bin.
Green open triangles show the luminosity function at $z=0.65-1.25$
calculated by Cowie et al.\ (2011) for their 5 NUV$<22$
pipeline extracted, optically confirmed, star-forming 
LAEs with rest-frame EW(\lya)$\ge 20$~\AA\ in multiple fields. 
Green solid
triangles show their luminosity function after they corrected
it to allow for sources with fainter continuum magnitudes.
Blue dotted curve shows the Schechter (1976) function
fit to the \lya\ luminosity function at $z=0.194-0.44$
from Cowie et al.\ (2010). Red solid curve shows the fit
at $z=3.1$ from Gronwall et al.\ (2007). Red dashed
curve shows the Gronwall et al.\ (2007) luminosity function
with its normalization ($\phi_{\star})$ reduced by a factor
of 30. This provides a reasonable match to the $z=0.67-1.16$
luminosity function.
\label{high_la_lumfun}
}
\end{inlinefigure}

This confirms what was previously reported by Cowie et al.\ (2011).
On the figure we also show the luminosity function they 
calculated for their 5 NUV$<22$ pipeline extracted, optically confirmed, 
star-forming LAE galaxies from multiple fields 
with rest-frame EW(\lya)$\ge 20$~\AA\
(green open triangles).
In addition, we show their luminosity 
function after they corrected it (green solid triangles)
to allow for sources with fainter
continuum magnitudes (see Deharveng et al.\ 2008 and Cowie et al.\ 2010 
for details).  Since many of the LAEs with \lya\ luminosities
$\sim 10^{43}$~erg~s$^{-1}$ have NUV$>22$, this
correction is substantial.  It is these corrected points that we 
should compare with the present luminosity function.  

The $z=0.67-1.16$ luminosity function 
presented here needs to be corrected for 
contamination from any remaining unidentified AGNs, 
which will slightly lower the
overall normalization. We postpone these corrections to a subsequent
paper, where we will present LAE catalogs determined
from data cube extractions of all the deepest {\em GALEX\/} grism 
fields, together with the optical spectroscopic follow-up necessary 
to remove the residual AGNs in the sample 
(I.~Wold et al.\ 2012, in preparation).
However, even allowing for this, we can see that there is a 
dramatic evolution in the luminosity function between $z\sim0.3$ and 
$z\sim0.9$ and that high-luminosity \lya\ sources seen at still 
higher redshifts are already present at $z\sim1$.

\section{Summary}
\label{summary}

In this paper, we described a new method for obtaining a flux-limited
sample of LAEs from {\em GALEX\/} grism data.  We showed
how for intensively observed fields with large numbers of independent
grism images we could construct three-dimensional (two spatial axes 
and one wavelength axis) data cubes from the grism images.
We then constructed the data cube for the central region of the
CDFS-00, the field with the deepest {\em GALEX\/} grism observations,
and used it to carry out a search for emission-line objects.

We detected 28 new $z\sim 1$ LAEs in our data cube search of the CDFS-00.
By comparing with the ECDFS X-ray data, existing optical 
spectroscopy, and deep $U$-band imaging,
we determined that the bulk of our new LAEs are 
real and that the UV spectroscopic redshifts based on the
\lya\ identifications are reliable.
In looking at the high-resolution {\em HST\/} images that are 
available for some of the LAEs in our new sample, as well as
for some of the LAEs from the {\em GALEX\/} pipeline sample,
we found that the sources
classified as AGNs are dominated by compact nuclei, while the
sources that appear to be star-forming galaxies are primarily resolved
small blue galaxies, often with multiple components. 
One of our new LAEs is spatially resolved with a linear
major-axis diameter of 120~kpc.  This places it in the category
of the giant \lya\ blobs, though at the low end in terms of
its \lya\ luminosity.

The LAEs that had previously been found by the {\em GALEX\/} 
pipeline were limited by the pipeline magnitude limit of NUV=22, 
but the new LAEs extend to much fainter magnitudes.  Moreover, 
they also have higher EWs, since for a
given \lya\ flux and redshift, the NUV=22 continuum cut corresponds
to an EW, which becomes progressively lower
as one moves to lower \lya\ fluxes.
The consequences of this are severe: the pipeline
progressively misses more sources from a \lya\ flux of 
$\sim8\times 10^{-15}$~erg~cm$^{-2}$~s$^{-1}$ down to a \lya\ flux of
$\sim2\times10^{-15}$~erg~cm$^{-2}$~s$^{-1}$, at which point it 
misses nearly all of them.  

Finally, we computed the CDFS-00 LAE galaxy luminosity function for 
the redshift interval $z=0.67-1.16$ using both the very small sample 
of LAE galaxies from the pipeline sample of
Cowie et al.\ (2011; 2 galaxies) and the larger sample of LAE 
galaxies from our data cube search.
We confirmed a dramatic evolution in the luminosity function 
between $z\sim 0.3$ and $z\sim 1$.

In the future, we plan to undertake data cube searches on the 
other deep {\em GALEX\/} fields to obtain a much larger sample of
$z\sim1$ LAEs (I.~Wold et al.\ 2012, in preparation).
This will enable us to make a precise determination
of the luminosity function at this redshift. We also plan to do
extensive optical spectroscopic follow-up to better assess the 
level of AGN contamination in the $z\sim 1$ LAEs.

\acknowledgements

We thank the referee for providing an interesting and thoughtful
report that helped us to improve the paper. We thank Michael
Cooper for supplying spectra from the Arizona CDFS
Environment Survey (ACES).
We gratefully acknowledge support from the University of Wisconsin
Research Committee with funds granted by the Wisconsin Alumni
Research Foundation and the David and Lucile Packard Foundation
(A.~J.~B.), as well as from NSF grant AST-0709356 (L.~L.~C.).


\newpage

\begin{deluxetable}{ccccccccccccc}
\renewcommand\baselinestretch{1.0}
\tablewidth{0pt}
\tablecaption{Emission Line Sample}
\scriptsize
\tablehead{Name & R.A. & Decl. & NUV & FUV & $z_{\rm galex}$ & $f$ & $\log L$ & EW & UV & $\log f$ & $z_{\rm ground}$ & Opt \\ & (J2000.0) & (J2000.0) &
(AB) & (AB) & & (Ly$\alpha$) & (Ly$\alpha$) & (Ly$\alpha$) &  Type & $(2-8~{\rm keV})$ & & Type\tablenotemark{a} \\ 
& & & & & & (erg/cm$^2$/s) & (erg/s) & (\AA) & & (erg/cm$^2$/s) & & \\
(1) & (2) & (3) & (4)  & (5) & (6) & (7) & (8) & (9) & (10) & (11) & (12) &(13) }
\startdata
GALEX033112-281517    &  52.800373   & -28.254972   &  23.19   &  27.56   &    1.160 &   77$\pm$3 &         43.16 &     109$\pm$2 &     \nodata &         \nodata &         \nodata &          \cr
GALEX033359-275326    &  53.499123   & -27.890778   &  22.14   &  24.11   &    1.144 &   24$\pm$3 &         43.45 &      82$\pm$4 &         AGN &         \nodata &         \nodata &          \cr
GALEX033301-273227    &  53.255127   & -27.540943   &  23.64   & -27.39   &    1.137 &   20$\pm$3 &         42.89 &      92$\pm$6 &     \nodata &         \nodata &         \nodata &          \cr
GALEX033146-272942    &  52.942501   & -27.495249   &  22.81   &  25.06   &    1.133 &   15$\pm$3 &         42.71 &      28$\pm$8 &     \nodata &         \nodata &         \nodata &          \cr
GALEX033230-280804    &  53.125668   & -28.134527   &  22.42   &  23.76   &    1.126 &   18$\pm$3 &         43.11 &      50$\pm$6 &         AGN &         \nodata &         \nodata &          \cr
GALEX033348-280216    &  53.453625   & -28.037998   &  23.49   &  25.57   &    1.125 &   10$\pm$3 &         42.93 &      89$\pm$5 &     \nodata &         \nodata &         \nodata &          \cr
GALEX033120-274917    &  52.837460   & -27.821417   &  23.00   & -29.55   &    1.116 &   13$\pm$2 &         42.76 &      38$\pm$5 &     \nodata &         \nodata &         \nodata &          \cr
GALEX033255-281320    &  53.231544   & -28.222471   &  22.51   &  28.98   &    1.084 &   14$\pm$3 &         42.80 &      28$\pm$5 &     \nodata &         \nodata &         \nodata &          \cr
GALEX033200-274319    &  53.001377   & -27.722055   &  22.45   &  24.24   &    1.042 &   10$\pm$3 &         43.11 &      61$\pm$7 &     \nodata &           -14.2 &          1.037\tablenotemark{b} &    BLAGN \cr
GALEX033228-273614    &  53.118168   & -27.604000   &  23.69   &  25.77   &    1.033 &   41$\pm$2 &         42.59 &      58$\pm$6 &     \nodata &           ECDFS &         \nodata &          \cr
GALEX033113-274949    &  52.807587   & -27.830278   &  22.54   & -28.61   &    1.032 &   16$\pm$2 &         42.70 &      26$\pm$5 &     \nodata &         \nodata &         \nodata &          \cr
GALEX033202-280320    &  53.010712   & -28.055723   &  23.55   &  25.24   &    1.017 &   26$\pm$2 &         42.89 &     104$\pm$3 &         AGN &           -13.7 &          1.014\tablenotemark{c} &     Sy2 \cr
GALEX033111-275506    &  52.798542   & -27.918472   &  24.00   & -27.28   &    0.996 &   15$\pm$2 &         42.62 &      89$\pm$8 &     \nodata &         \nodata &         \nodata &          \cr
GALEX033336-274224    &  53.402752   & -27.706667   &  23.04   &  26.32   &    0.978 &   28$\pm$2 &         43.10 &     116$\pm$9 &         AGN &           -14.7 &         \nodata &          \cr
GALEX033246-274154\tablenotemark{d}    &  53.195625   & -27.698416   &  23.41   &  25.82   &    0.977 &   14$\pm$2 &         42.86 &      93$\pm$3 &     \nodata &           ECDFS &         \nodata &          \cr
GALEX033204-273725    &  53.017002   & -27.623806   &  22.87   &  24.31   &    0.976 &    8$\pm$3 &         43.14 &     107$\pm$8 &         AGN &           -13.5 &         0.970\tablenotemark{e} &    BLAGN \cr
GALEX033335-273934    &  53.398125   & -27.659584   &  22.62   &  23.31   &    0.949 &    9$\pm$2 &         42.85 &      47$\pm$6 &     \nodata &           -14.2 &         \nodata &          \cr
GALEX033329-280127    &  53.373249   & -28.024277   &  23.23   &  25.71   &    0.941 &   23$\pm$2 &         42.62 &      48$\pm$1 &     \nodata &           ECDFS &         \nodata &          \cr
GALEX033206-281408    &  53.026459   & -28.235723   &  21.88   &  25.76   &    0.913 &   10$\pm$2 &         43.22 &      60$\pm$3 &     \nodata &         \nodata &         \nodata &          \cr
GALEX033044-280237    &  52.686874   & -28.043751   &  23.33   &  27.41   &    0.872 &   12$\pm$2 &         42.57 &      56$\pm$7 &     \nodata &         \nodata &         0.855 & NELG       \cr
GALEX033251-280809    &  53.214336   & -28.135973   &  22.76   &  26.13   &    0.836 &   39$\pm$2 &         42.67 &      44$\pm$0 &     \nodata &         \nodata &         0.832 &  NELG         \cr
GALEX033045-274506    &  52.689747   & -27.751862   &  23.21   &  25.77   &    0.830 &    7$\pm$2 &         42.64 &      65$\pm$2 &     \nodata &         \nodata &         \nodata &          \cr
GALEX033314-274834    &  53.310581   & -27.809526   &  24.24   &  25.30   &    0.830 &   11$\pm$2 &         42.49 &     119$\pm$3 &     \nodata &           ECDFS &         \nodata &          \cr
GALEX033057-273316    &  52.739708   & -27.554583   &  22.49   &  24.02   &    0.795 &   20$\pm$2 &         42.72 &      43$\pm$6 &     \nodata &         \nodata &         0.790 &         NELG  \cr
GALEX033124-275625    &  52.851665   & -27.940474   &  22.78   &  25.39   &    0.775 &    9$\pm$2 &         42.62 &      48$\pm$9 &     \nodata &           ECDFS &         0.776\tablenotemark{f}  & NELG    \cr
GALEX033235-274059    &  53.148876   & -27.683195   &  23.74   &  25.95   &    0.746 &    7$\pm$2 &         42.70 &     150$\pm$9 &     \nodata &           ECDFS &         0.735\tablenotemark{g} &    NELG \cr
GALEX033146-274846    &  52.942459   & -27.812805   &  23.18   &  25.48   &    0.742 &   19$\pm$2 &         42.78 &     107$\pm$3 &     \nodata &           ECDFS &         \nodata &          \cr
GALEX033131-273429    &  52.880333   & -27.574804   &  22.57   &  25.54   &    0.695 &    8$\pm$1 &         43.21 &     191$\pm$8 &     \nodata &           -14.1 &         0.688\tablenotemark{h} &    Sy2 \cr
\enddata
\label{faint_z1_sample_table}
\tablenotetext{a}{BLAGN = broad-line AGN, i.e., presence of at least one emission line having FWHM$>2000$~km~s$^{-1}$;
Sy2 = Seyfert 2, i.e., [NeV] emission; NELG = narrow emission-line galaxy, 
i.e., at least one emission line and no BLAGN signatures or [NeV] emission}
\tablenotetext{b}{Szokoly et al.\ (2004)}
\tablenotetext{c}{Treister et al.\ (2009)}
\tablenotetext{d}{Giant \lya\ blob; see Figures~\ref{blob} and \ref{new_blob}}
\tablenotetext{e}{Treister et al.\ (2009); Szokoly et al.\ (2004) give $z=0.977$}
\tablenotetext{f}{Cooper et al.\ (2011)}
\tablenotetext{g}{Vanzella et al.\ (2008)}
\tablenotetext{h}{Silverman et al.\ (2010)}
\end{deluxetable}

\end{document}